\documentclass[twocolumn,tighten,times,twocolappendix]{aastex631}

\usepackage{xspace}

\usepackage{comment}
\usepackage{amsmath}
\usepackage{CJKutf8}
\defcitealias{Ritter88}{R88}
\defcitealias{Fuller24}{FT24}

\begin{document}

%\title{White dwarfs with transiting planetary debris in the Zwicky Transient Facility: Six new candidates and a unique metric space to identify them}

%\title{From Rapid Winkers to Slow Blinkers and a Gas Disk: Six New White Dwarfs with Transiting Planetary Debris from the Zwicky Transient Facility Identified in a Novel Variability Metric Space}
%\title{Debris Transits in White Dwarfs, Fast and Slow: Six New Candidates, one with Gas Disk, from ZTF Identified in a Novel Metric Space}
%\title{Abell 35 and Its Types in Context: Post-Mass Transfer Inflated Stars, and Precursors to Blue Lurkers and Wide WD+MS Binaries?}
\title{Enhanced Mass Transfer from Extended Shock-Supported Stellar Chromospheres}
\shorttitle{Chromospheric Mass Transfer}
\shortauthors{Bhattacharjee et al.}

\correspondingauthor{Soumyadeep Bhattacharjee}
\email{sbhatta2@caltech.edu}

\author[0000-0003-2071-2956]{Soumyadeep Bhattacharjee}
\affiliation{Department of Astronomy, California Institute of Technology, 1216 E. California Blvd, Pasadena, CA, 91125, USA}

\author[0000-0002-4544-0750]{Jim Fuller}
\affiliation{TAPIR, Mailcode 350-17, California Institute of Technology, Pasadena, CA 91125}

\author[0000-0002-6871-1752]{Kareem El-Badry}
\affiliation{Department of Astronomy, California Institute of Technology, 1216 E. California Blvd, Pasadena, CA, 91125, USA}

%% ABSTRACT %%
\begin{abstract}

Mass transfer (MT) in binary systems is a crucial and complex process. Even before the donor's photosphere fills its Roche lobe, matter from its outer atmosphere can reach the inner Lagrange point, thus initiating interaction in the so-called optically thin phase. Most existing models account for a very tenuous atmosphere supported by thermal gas pressure, thus predicting very little MT until the donor fills its Roche lobe. However, recent hydrodynamic simulations of giant stars suggest that shock-driven outflows can create substantially thicker ``chromospheres'' extending to a few stellar radii. In this work, we derive a simple analytic model for a shock-supported chromosphere in a binary potential and calculate the corresponding MT rate. We also implement our prescription in the stellar evolution code MESA. We show that chromospheric MT rates can be orders of magnitude higher than prior estimates for optically thin MT, especially for stars high on the red giant or asymptotic giant branch. This allows for substantial accretion before Roche lobe overflow, both for systems that undergo stable mass transfer and for systems that go through common envelope evolution. Thus, for a given initial binary configuration, this may alter the final period and component masses compared to those predicted classically. Our model is similar to Wind Roche-lobe Overflow, but it is physically rather than empirically motivated. We discuss implications for various post-interaction binaries that can be tested with population synthesis studies and observed binary populations. Additionally, we have identified an inaccuracy in the current implementation of the optically-thin \cite{Ritter88} prescription in MESA, which we correct. 

%We find that the resultant MT rates can be orders of magnitude higher than the classical estimates from thermal atmospheres, especially at lower Roche filling factors of the donors. We implement our prescription in the stellar evolution code MESA to show that such enhanced interaction can substantially alter the fate of the binary compared to our current understanding by triggering interaction at much wider orbits, enhancing donor stripping, and transferring up to orders of magnitude more mass to the companion. This can re-shape our interpretation of the various populations of AU-scale post-interaction binaries. The chromosphere also allows for substantial optically thin MT in binaries that eventually undergo common envelope evolution (CEE) in the optically thick phase, potentially explaining the observed accretion signatures in several post-CEE systems. Compared to the existing wide-separation MT prescription of Wind Roche-lobe Overflow, our model is more physically motivated and fundamentally derived. Future population synthetic studies and comparison with numerical simulations will further test and calibrate our model.

\end{abstract}

%% Keywords should appear after the \end{abstract} command. 
%% The AAS Journals now uses Unified Astronomy Thesaurus concepts:
%% https://astrothesaurus.org
% https://astrothesaurus.org/thesaurus/alphabetical-browse/
\keywords{Binary Stars (154) -- Interacting binary stars (801) -- Stellar Accretion (1578) -- Roche lobe overflow (2155) -- Stellar chromospheres (230)}

\section{Introduction}\label{sec:intro}

Mass transfer is a crucial phenomenon in binary systems which can dramatically alter the evolution of the stars \citep{Mathieu25, vanSon26}. This occurs when the matter in the outer layers of one of the stars, called the donor, comes close to the inner Lagrange point (L1) of the binary system. This matter gets channeled through the L1 nozzle and accreted onto the companion star, called the accretor. Mass transfer is often triggered as the donor star evolves off the main sequence, expands into a red giant, and starts to fill its Roche lobe. This reduces the potential difference between the stellar photosphere and the L1 point, thus enabling the flow of matter from the donor star to the nozzle.

The outcomes of mass transfer can be diverse. The process can be stable, where the donor stays mostly contained within its Roche lobe. On the other hand, mass transfer from highly evolved convective donors, like asymptotic giant branch (AGB) or red-supergiant (RSG) stars, often leads to instability and common envelope (CE) evolution (CEE) \citep{Ivanova13}. This leads to substantial mass ejection and orbital shrinkage. The products include systems like hot subdwarfs (e.g. \citealt{Han03}), short-period binaries with a white dwarf (WD, e.g. \citealt{Rebassa-Mansergas07, Sun18, Li19, Jones20}), cataclysmic variables (e.g. \citealt{Meyer79}), and double compact objects \citep{Vigna-Gomez20}.

%The products usually are relatively wide binaries with orbital periods $\gtrsim$$10^2$~days and include systems like hot subdwarfs (e.g. \citealt{Han03}), low-mass red giants (e.g. \citealt{El-Badry22,Jayasinghe22}), and helium white dwarfs (e.g. \citealt{Sun18,Li19}). This typically results in short-period ($\lesssim$$1$~day) orbits. Short-period binaries with a white dwarf (WD, e.g. \citealt{Rebassa-Mansergas07, Jones20}), cataclysmic variables (e.g. \citealt{Meyer79}), and double compact objects \citep{Vigna-Gomez20} form via CEE.

The range of possible outcomes has recently been substantially expanded with the discovery of a population of post-mass transfer AU-scale (orbital periods of $P_{\rm orb}\sim10^2-10^3$~days) binaries, consisting of a main-sequence star and a post-AGB white dwarf (WD, \citealt{Shahaf23,Yamaguchi24b,Yamaguchi25}) or a post-RSG neutron star (NS, \citealt{Elbadry24ns}). Determining the formation channels of these systems is an area of active research, but they certainly challenge the classically assumed boundary between stable and unstable mass transfer.

Disentangling this zoo of mass transfer outcomes depends crucially on our understanding of the mass transfer mechanism and the underlying physics. The foundation was laid by \cite{Lubow75}, who argued that the gas attains sonic speed at L1 and gets ballistically transferred via the nozzle. This was then used by \cite[][henceforth, \citetalias{Ritter88}]{Ritter88} and then by \cite{Kolb90} to develop a complete prescription for the mass transfer rate, $\dot{M}$, for a Roche lobe filling star in a binary. The former focuses on the optically thin regime, where the donor's photosphere is still under-filling the Roche lobe, but the mass from its outer atmosphere reaches L1 and gets transferred. The latter work extends it to photospheric Roche lobe overflow, where an adiabatic equation of state was used instead. These prescriptions are widely used in modeling mass transfer outcomes with stellar evolution codes like MESA \citep{Paxton15} and population synthesis codes like POSYDON \citep{Fragos23}. 

In this work, we shall focus on the optically thin regime of mass transfer. The primary assumption in \citetalias{Ritter88} is that the stellar atmosphere is supported by the thermal pressure of an isothermal atmosphere. Such atmospheres are usually very thin, with scale heights $\lesssim$$10^{-2}\times$ the stellar radius. The optically thin mass transfer phase is thus short-lived in this formalism, with a sharp transition to the optically thick phase. 

In recent times, several modifications to the \citetalias{Ritter88} prescription have been proposed. For example, \cite{Cehula23} performed a careful derivation of the hydrodynamics of the L1 flow (in contrast to the hydrostatic atmosphere assumed in \citetalias{Ritter88}) and experimented with various equations of state (EOS). More recently, \cite{Ivanova24} combined the optically thin and thick phases into a uniform formalism and also experimented with five different atmospheric models. Both works found $\dot{M}$ to differ by factors of $\approx$$2-4$ from the \citetalias{Ritter88} prescription, depending on the atmospheric model assumed. Nevertheless, all the works share the common assumption that the stellar atmosphere is thermally supported, and is thus thin, comparable to the pressure scale height at the stellar surface. 

Numerical simulations of giant stars, however, suggest otherwise. A combination of coherent stellar pulsations and stochastic convection drives shock waves which elevate matter above the stellar photosphere. This expands the stellar atmosphere \citep{Hofner03} and creates a dense ``chromosphere'' of gravitationally bound, optically thin material above the photosphere. Several subsequent numerical hydrodynamical simulations of AGB and RSG stars have confirmed this scenario \citep[for example,][]{Freytag17,Freytag23,Ma25}. More recently, \cite[][henceforth, \citetalias{Fuller24}]{Fuller24} derived an analytic hydrostatic formalism for a chromosphere arising from convection-driven shock waves. The chromospheres in all these studies are much denser and more extended than thermally supported atmospheres.

This is expected to significantly affect binary interaction. The extended chromosphere can reach the L1 point at much lower photospheric Roche filling factors of the donor. This will greatly enhance the mass transfer rate compared to that given by \citetalias{Ritter88}, potentially modifying the outcome. Consequently, the binary components can interact at larger separations compared to those allowed by the classical \citetalias{Ritter88} prescription -- two stars not supposed to interact in the classical formalism can exchange a substantial amount of mass when the extended chromospheres are taken into account. 

There is observational evidence on multiple fronts already suggesting enhanced mass transfer at low Roche filling factors. For example, in several known actively mass transferring systems (e.g., symbiotic systems), the $\dot{M}$ values suggested by observations are much larger than those expected from the classical prescriptions at the donors' Roche fill factor \citep{Mikoajewska12, Gaudin26}. Several post-mass transfer binaries, e.g., Blue Lurkers \citep{Leiner19} and Blue Straggler Stars \citep{Mathieu09}, have long orbital periods where the donors could not have filled their Roche lobes, even in their giant phases. Other post-mass transfer systems such as Barium stars suggest total transferred masses much in excess of what \citetalias{Ritter88}-like prescriptions can supply \citep{Sarmah26}. Chromospheric mass transfer can potentially address these observational puzzles.

%Very recently, \cite{Sarmah26} studied the population of Ba stars to estimate the total transferred masses to be $\approx$$0.1-0.5~M_{\odot}$. Given that their orbital periods lie in the range of $\sim$$10^3-10^4$~days, the classical prescription would fail to account for the mass transferred. 

In this work, we present our calculation for the mass transfer from extended stellar chromospheres. We focus on the region within the dust formation radius and do not include the calculations for wind-driven accretion from beyond (which shall be presented separately). We apply our results mostly to AGB stars, but our calculations are also valid for other types of red giants and RSGs, and we discuss the implications for massive star binaries wherever possible.

The rest of the paper is as follows. In Section~\ref{sec:conceptual_overview} we present a brief overview of both the classical \citetalias{Ritter88} formalism and chromospheric mass transfer to highlight the basic differences. In Section~\ref{sec:chrom_dyn} we present the detailed derivation of the chromospheric density profile and mass transfer rates. We implement our calculations in MESA and present results for three representative binaries in Section~\ref{sec:mesa_chrom}. In Section~\ref{sec:discussion} we discuss the possible implications and future observational/theoretical tests. Finally, in Section~\ref{sec:conclusions} we present our conclusions.

\section{Conceptual Overview}\label{sec:conceptual_overview}

Before delving into the details of the chromospheric mass transfer, we provide a general conceptual overview of optically thin mass transfer and how it manifests in both the classical \citetalias{Ritter88} prescription and chromospheric mass transfer. The general equation governing the mass transfer rate through the L1 Lagrange nozzle is given by:
\begin{equation}\label{eq:mdot_basic}
\begin{split}
    \dot{M} = \int\rho(\phi)\,v(\phi)\,dA_N(\phi)=\int\rho\,v\,\frac{dA_N}{d\phi}\,d\phi
\end{split}
\end{equation}
where $\rho$ and $v$ denote the density and velocity, and $A_N$ denotes the nozzle area -- all being functions of the potential, $\phi$. Physically, $A_N(\phi)$ denotes the cross-sectional area perpendicular to the binary axis at L1 bounded by the corresponding equipotential surface. To first order in $\Delta\phi = \phi-\phi_{L1}$, the mass transfer rate takes the approximate form:
\begin{equation}\label{eq:oom_mdot_l1}
    \dot{M}\sim\rho_{\rm L1}v_{\rm L1}\left|\frac{dA_{\rm N}}{d\phi}\right|_{\rm L1}\Delta\phi\\\sim\rho_{\rm L1}v_{\rm L1}^3\left|\frac{dA_{\rm N}}{d\phi}\right|_{\rm L1},
\end{equation}
where the last approximation arises due to the kinetic energy setting the potential window through which matter gets transferred through L1, i.e., $\Delta\phi\sim v_{\rm L1}^2$. Additionally, one can show that:
\begin{equation}\label{eq:oom_dadphi_l1}
    \left|\frac{dA_{\rm N}}{d\phi}\right|_{\rm L1}\sim\frac{P_{\rm orb}^2}{4\pi},
\end{equation}
where $P_{\rm orb}$ is the orbital period (see Section \ref{subsec:chrom_mdot}).

In the optically thin regime, the stellar photosphere is still within its Roche lobe. The matter at L1 is thus provided by the ``atmosphere'' of the star extending beyond its photosphere. The classically assumed stellar atmosphere is formed from the thermal pressure of the photospheric gas, which forms the basis of the \citetalias{Ritter88} prescription. On the other hand, when the pressure support is provided by convection-driven photospheric shocks, we call it the shock-supported chromosphere.

\begin{comment}
When the gas motion giving rise to this atmosphere has a characteristic speed, $v_c$, constant in space and time, hydrostatic equilibrium equation yields an exponential dependence of the density on the potential difference from the stellar surface. Barring model-dependent sub-dominant pre-factors, the density profile roughly goes as:
\begin{equation}\label{eq:oom_dens_l1}
    \rho_{\rm L1}\sim\rho_{\rm ph}e^{-\frac{\phi_{\rm L1}-\phi_{\rm ph}}{v_c^2}},
\end{equation}
where $\phi$ denotes the potential. The characteristic speed of the gas particles sets the scale height of this atmosphere. For the purpose of this overview, we work in the regime of $\phi_{\rm ph}<<\phi_{\rm L1}$ where the difference between the two prescriptions is the most prominent. In addition, it enables the simplification of the equations and eases the introduction of the key parameters. We then have: 
\begin{equation}\label{eq:phil1_sub_phiph_eq_vesc2}
    \phi_{\rm L1}-\phi_{\rm ph}\approx|\phi_{\rm ph}|=v_{\rm esc}^2/2,
\end{equation}
where
\begin{equation}
    v_{\rm esc}=\sqrt{2GM/R_{\rm ph}}
\end{equation}
is the escape velocity at the stellar surface. 
\end{comment}

\subsection{The classical prescription}
%The detailed derivation can be found in the original work by \cite{Ritter88} and several recent works (quite unnecessarily) revisiting this prescription (for example, \citealt{Marchant21}). We instead provide here a brief conceptual overview and introduce the essential parameters. The optically thin mass transfer process focuses on the regime where the  stellar photosphere is still within its Roche lobe. We now focus on the regime where the stellar photosphere is still within its Roche lobe -- the optically thin regime of mass transfer. 

The photospheric isothermal sound speed is given by $c_T=\sqrt{k_BT_{\rm ph}/\mu}$, where $T_{\rm ph}$ is the photospheric temperature and $\mu$ is the mean molecular weight. Hydrostatic equilibrium maintained by a gas with thermal pressure $P=\rho c_T^2$ yields an exponential density profile given by:
\begin{equation}\label{eq:oom_dens_l1}
    \rho_{\rm L1}\sim\rho_{\rm ph}e^{-\frac{\Delta\phi}{c_T^2}},
\end{equation}
where $\Delta\phi=\phi_{\rm L1}-\phi_{\rm ph}$ is the difference in gravitational potential between the photosphere and the L1 point. When the star under-fills the Roche lobe substantially, we have $\Delta\phi\approx|\phi_{\rm ph}|=v_{\rm esc}^2/2$, where $v_{\rm esc} = \sqrt{2 G M/R_{\rm ph}}$ is the escape speed at the stellar photosphere of radius $R_{\rm ph}$. On the other hand, when Roche filling, we have $\Delta\phi\approx0$. Thus, for convenience, we write $\Delta\phi=\mathcal{F}^2v_{\rm esc}^2/2$, where $0\lesssim\mathcal{F}\lesssim1$ depending on the Roche filling factor of the donor. Now, the assumption that the gas motions are thermalized to $T_{\rm ph}$ throughout the atmosphere sets $v_{\rm L1}=c_T$ \citep{Lubow75}. Using Equations~\ref{eq:oom_mdot_l1},~\ref{eq:oom_dadphi_l1}, and \ref{eq:oom_dens_l1}, the mass transfer rate is then roughly given by:
\begin{equation}\label{eq:oom_dens_l1_approx_fT}
    \dot{M}_R\sim\rho_{\rm ph}v_{\rm esc}^3f_T^3\frac{P_{\rm orb}^2}{4\pi}e^{-\frac{\mathcal{F}^2}{2f_{T}^2}},
\end{equation}
where we have defined the velocity parameter as:
\begin{equation}\label{eq:fT}
    f_T=c_T/v_{\rm esc}.
\end{equation}
For most stars, $f_T\in[0.01,~0.1]$. This severely suppresses the optically thin mass transfer rate unless the star sufficiently fills its Roche lobe such that $\mathcal{F}\approx0$.

%\citetalias{Ritter88} assumes that the stellar atmosphere is supported by the thermal motion of the particles at the photosphere, set by the photospheric temperature, $T_{\rm ph}$. Thus, the characteristic gas speed becomes the photospheric thermal speed: $v_c=c_T=\sqrt{k_BT_{\rm ph}/\mu}$. Additionally, they also assume that the gas motions are thermalized to $T_{\rm ph}$ throughout the atmosphere, which also sets the mass transfer velocity at the L1 point. This yields $v_{\rm L1}=c_T$. 

\subsection{Shock-supported chromosphere}

The formation of the chromosphere, which is supported by momentum from shocks, is fundamentally different from that of the thermal atmosphere. However, under the simplifying assumptions of our model detailed in Section~\ref{subsec:chrom_dens_profile}, the density profile broadly retains the exponential form of Equation~\ref{eq:oom_dens_l1} with the sole modification of replacing $c_T$ with the shock speed, $v_s$. Thus, the density at L1 is roughly given by:
\begin{equation}\label{eq:oom_dens_l1_chrom}
    \rho_{\rm L1}\sim\rho_{\rm ph}e^{-\frac{\mathcal{F}^2v_{\rm esc}^2}{2v_s^2}}.
\end{equation}
Since the shocks are launched by convective motions in stellar envelopes, the shock speed varies probabilistically. \citetalias{Fuller24} assumes a Gaussian distribution of the shock velocities:
\begin{equation}\label{eq:vs_gauss}
    p(v_s)=\sqrt{\frac{2}{\pi}}\frac{1}{v_{\rm con}}e^{-\frac{v_s^2}{2v_{\rm con}^2}},
\end{equation}
where $v_{\rm con}$ is the root-mean-square convective speed. Thus, to obtain the mean density at L1, Equation~\ref{eq:oom_dens_l1_chrom} needs to be convolved with Equation~\ref{eq:vs_gauss}. Together with Equation~\ref{eq:oom_mdot_l1}, the mass transfer rate thus becomes:
\begin{equation}\label{eq:oom_chrom_mdot}
\begin{split}
    \dot{M}_c\sim\frac{\rho_{\rm ph}}{v_{\rm con}}\frac{P_{\rm orb}^2}{4\pi}\int_0^{\infty}v_s^3 e^{-\left(\frac{\mathcal{F}^2v_{\rm esc}^2}{2v_s^2}+\frac{v_s^2}{2v_{\rm con}^2}\right)}\,dv_s.
\end{split}
\end{equation}
We now introduce a parameter analogous to $f_T$ for convective motion:
\begin{equation}\label{eq:fcon}
    f_{\rm con} = v_{\rm con}/v_{\rm esc}.
\end{equation}
For the majority of cases, the peak shock velocity can be approximated as $v_s^p\sim\sqrt{v_{\rm esc}v_{\rm con}}=v_{\rm esc}f_{\rm con}^{1/2}$. Thus, the mass transfer rate can be approximated as:
\begin{equation}\label{eq:oom_chrom_mdot_approx}
\begin{split}
    \dot{M}_c\sim\frac{\rho_{\rm ph}}{v_{\rm con}}\frac{P_{\rm orb}^2}{4\pi}(v_s^p)^3\int_0^{\infty} e^{-\left(\frac{\mathcal{F}^2v_{\rm esc}^2}{2v_s^2}+\frac{v_s^2}{2v_{\rm con}^2}\right)}\,dv_s\\\sim\rho_{\rm ph}v_{\rm esc}^3f_{\rm con}^{3/2}\frac{P_{\rm orb}^2}{4\pi}e^{-\frac{\mathcal{F}}{f_{\rm con}}}
\end{split}
\end{equation}

%The integrand in the above equation peaks at a shock speed value of $v_s^p\sim v_{\rm esc}f_{\rm con}\times g\left(\frac{F}{f_{\rm con}}\right)$, where $g(x)$ takes a value roughly between $1~(x\rightarrow0)$ and $x^{3/2}~(\text{large}~x)$. For most values of $F/f_{\rm con}$, this is expected to be order unity. 

\begin{figure}
    \centering
    \includegraphics[width=1\linewidth]{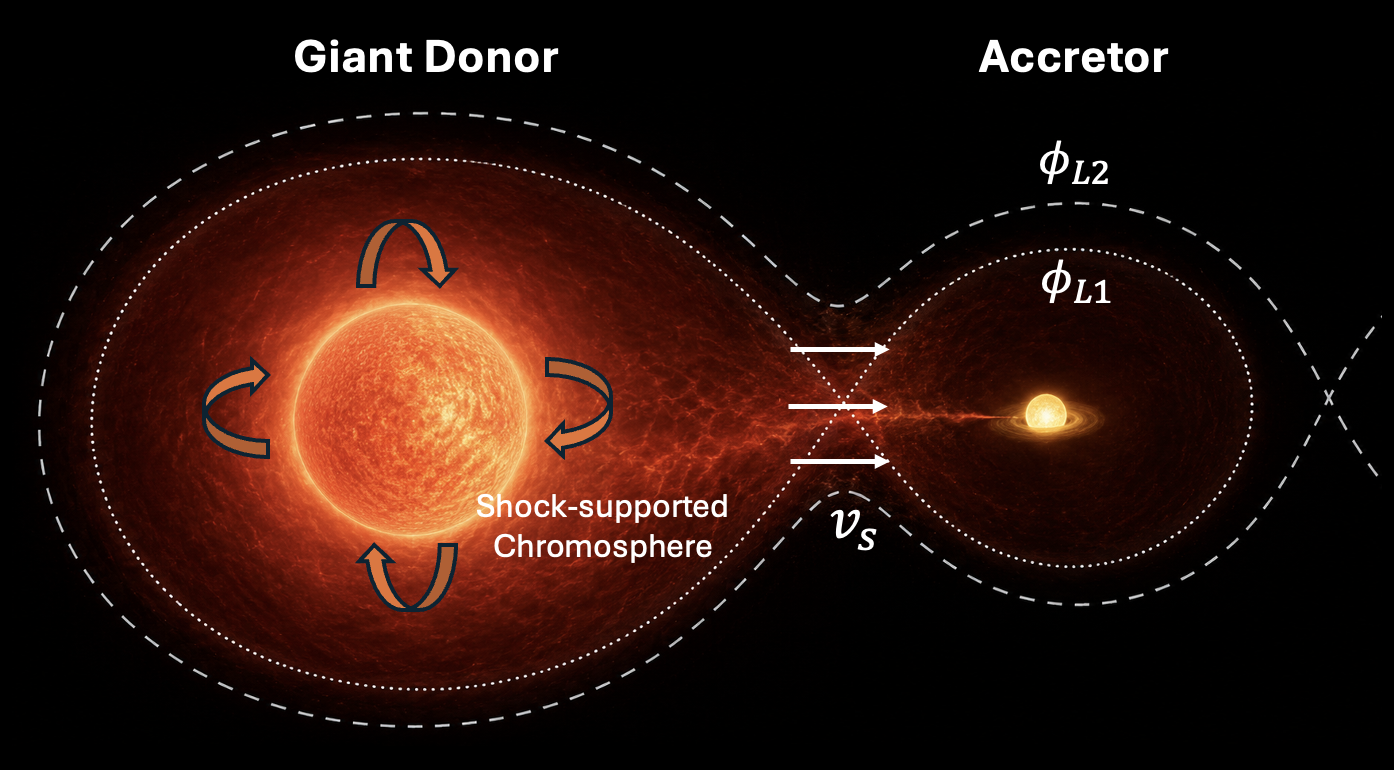}
    \caption{Cartoon demonstrating chromospheric mass transfer in a binary with $M_d>M_a$. The giant is under-filling its Roche lobe (dotted line, the L1 equipotential) but the shock-supported chromosphere extends to larger potentials which enables mass transfer. The chromospheric gas between the L1 and L2 (dashed line) potentials gets transferred to the accretor. The characteristic mass transfer velocity is the shock speed, $v_s$.}
    \label{fig:cool_figure}
\end{figure}

%\sim\rho_{\rm ph}\bar{v}_s^3\frac{P_{\rm orb}^2}{4\pi}e^{-\frac{\mathcal{F}}{f_{\rm con}}}
\begin{comment}
This yields:
\begin{equation}\label{eq:mean_chrom_rho_l1}
\begin{split}
    \bar{\rho}_{\rm L1}\sim\frac{\rho_{\rm ph}}{v_{\rm con}}\int_0^{\infty} e^{-\left(\frac{\mathcal{F}^2v_{\rm esc}^2}{2v_s^2}+\frac{v_s^2}{2v_{\rm con}^2}\right)}\,dv_s\\\sim\rho_{\rm ph}e^{-\frac{\mathcal{F}}{f_{\rm con}}},
\end{split}
\end{equation}
where we have introduced a parameter analogous to $f_T$ for convective motion:
\begin{equation}\label{eq:fcon}
    f_{\rm con} = v_{\rm con}/v_{\rm esc}.
\end{equation}
Note that the integrand in Equation~\ref{eq:mean_chrom_rho_l1} peaks at a velocity $\sqrt{\mathcal{F}v_{\rm con}v_{\rm esc}}$, which thus becomes the characteristic velocity for mass transfer. The chromospheric mass transfer rate thus roughly becomes:
\begin{equation}
    \dot{M}_c\sim\rho_{\rm ph}\mathcal{F}^{\frac{3}{2}}v_{\rm esc}^3f_{\rm con}^{\frac{3}{2}}\frac{P_{\rm orb}^2}{4\pi}e^{-\frac{\mathcal{F}}{f_{\rm con}}}.
\end{equation}
\end{comment}

Similar to $f_T$, $f_{\rm con}$ also takes values in the range of $[0.01,~0.1]$. However, compared to Equation~\ref{eq:oom_dens_l1_approx_fT}, the classical prescription depends on $1/f_T^2$ which is significantly steeper than $1/f_{\rm con}$, when the two parameters take values in a similar range. This difference is mainly brought about by the high-$v_s$ tail of the shock-speed distribution, which significantly increases the density at large radii compared to the thermally supported atmosphere. For a given Roche lobe filling factor, the chromosphere can thus yield a mass transfer rate orders of magnitude higher than the classical prescription. A diagrammatic representation of chromospheric mass transfer is shown in Figure~\ref{fig:cool_figure}.

\section{Deriving the Chromospheric Dynamics}\label{sec:chrom_dyn}

In the following sections, we present our detailed derivation of the chromospheric density profile and mass transfer rate in a binary potential. We focus on the region within the dust formation region where the shocked matter has not yet condensed to form dust, and has not yet been accelerated into an unbound wind.
%Thus, we use an ideal non-viscous isotropic fluid approximation for the chromosphere.

\subsection{Generalized Chromospheric density profile}\label{subsec:chrom_dens_profile}

We start with the Euler equation:
\begin{equation}\label{eq:euler}
    \rho\frac{\partial \vec{v}}{\partial t}+\rho\vec{v}\cdot\nabla\vec{v} = -\nabla P+\vec{f},
\end{equation}
where $\vec{v}$ is the velocity of the fluid, $P$ is the pressure, $\rho$ is the density and $\vec{f}$ represents any additional forces. For simplicity, we assume a steady flow in which the time-derivative vanishes. This is justified when the shock velocities discussed below are smaller than the escape speed and Equation~\ref{eq:euler} is averaged over a sufficiently long time. %Our problem of mass transfer, however, is intrinsically dynamic. Any $\dot{M}$ derived from hydrostatic approximation will only be valid as long as the mass transfer timescale is longer than the time it takes for subsequent shocks to replenish the transferred mass. We find that this is mostly valid, however, we recognize this approximation as a caveat of our proof-of-concept model. 

There are two important forces acting on the fluid, referring to \cite{Lubow75}. First, the gravitational and centrifugal force can be written as $\vec{f}_g=-\rho\nabla\phi$, where $\phi$ is the Roche potential. Second, the Coriolis force is $\vec{f}_c=-2\rho\Omega_{\rm orb}\hat{z}\times\vec{v}$, where $\hat{z}$ is the direction perpendicular to the orbital plane and $\Omega_{\rm orb}$ is the orbital angular speed. In our model, we include outgoing shocks as an additional force -- the main force supporting the chromosphere. We model this as a Reynolds stress: $\vec{f}_s=-\nabla\cdot(\rho\vec{v}_s\vec{v}_s)$. The total force is thus $\vec{f}=\vec{f}_g+\vec{f}_c+\vec{f}_s$. With this, and re-writing the advection term in terms of the Lamb vector, we get our final equation in the co-rotating frame as:
\begin{align}
    \label{eq:convective_momentum}
    &\frac{1}{2}\nabla v^2 + (\nabla\times\vec{v})\times\vec{v} +\frac{\nabla P}{\rho} + \nabla\phi \nonumber \\ & +2\Omega_{\rm orb}\hat{z}\times\vec{v} +\frac{1}{\rho}\nabla\cdot(\rho\vec{v}_s\vec{v}_s)=0.
\end{align}

We now aim to derive a Bernoulli-like quantity which is conserved along a streamline. Barring the shock term, the first five terms in the equation comprise the standard Euler equation in a rotating frame, worked on by \cite{Lubow75} -- taking a dot product with $\vec{v}$ yields the usual conserved Bernoulli parameter (their equations 6 and 7). Thus, it remains to express the shock term in Equation~\ref{eq:convective_momentum} as a total gradient -- to find a $Q$ such that
\begin{equation}\label{eq:vs_gradient}
\nabla\cdot(\rho\vec{v}_s\vec{v}_s)=\rho\nabla Q \, .
\end{equation}
There is, however, no simple solution in the three-dimensional Roche potential. However, under certain assumptions, it is possible to re-frame Equation~\ref{eq:convective_momentum} into a volume-averaged one-dimensional form which is sufficient for our simple model and enables ready implementation in one-dimensional codes like MESA.

The simplifying assumptions to facilitate this calculation are:
\begin{enumerate}

    \item The shock speed remains constant with radius. We follow the same overall justifications as in Section~2.1 of \citetalias{Fuller24}. Constant shock speeds are correct in the plane-parallel limit \citep{Bertschinger1985} and global simulations show approximately constant shock speeds \citep{Freytag17}.
    %-- isothermal shocks tend to accelerate down a decaying density gradient which counteracts the gravitational deceleration and maintains the constant speed \citep{Matzner99,Coughlin18}.

    \item The shock velocity is perpendicular to the equipotential surfaces. Horizontal components of the shocks do not support the fluid against gravity and are less important for determining chromospheric structure. Mathematically, we then have
    \begin{equation}\label{eq:vs}
        \vec{v}_s = v_s\hat{n},
    \end{equation}
    where $\hat{n}$ is the unit normal vector to a given equipotential surface.

    \item The density $\rho$ is only a function of the potential, $\phi$. Density variations on the equipotential surface can be neutralized through horizontal flows.
    
\end{enumerate}
None of these simplifications are expected to be exactly correct, but they are necessary to create a tractable model.

Neglecting any motion on the equipotential surface, we are interested only in fluxes out of equipotential volumes. We calculate this by inserting Equation~\ref{eq:vs} in Equation~\ref{eq:vs_gradient}, taking a dot product with $\hat{n}$, and integrating over the volume bounded by the equipotential surface. This yields for the left-hand side of Equation~\ref{eq:vs_gradient}:
\begin{equation}\label{eq:lhs_eval}
\begin{split}
    \int_{V_{\phi}}\hat{n}\cdot\nabla\cdot(\rho v_s^2\hat{n}\hat{n})~dV= \int_{V_{\phi}}n_i\partial_j(\rho v_s^2 n_i n_j)~dV \\ =\int_{V_{\phi}}\left[n_i n_i\partial_j(v_s^2\rho n_j) + \frac{1}{2}v_s^2\rho n_j \partial_j( n_in_i)\right]~dV \\ =  \int_{V_{\phi}}\nabla\cdot(\rho v_s^2\hat{n})~dV=v_s^2\rho(\phi)A_{\phi} \, ,
\end{split}
\end{equation}
where $A_\phi$ is the area of the equipotential surface.
The right-hand side evaluates to:
\begin{equation}\label{eq:rhs_eval}
\begin{split}
    \int_{V_{\phi}}\hat{n}\cdot\rho\nabla Q~dV=\oint_{A_{\phi}}\int_{\phi}\frac{\nabla\phi}{|\nabla\phi|}\cdot\rho\frac{dQ}{d\phi}\nabla\phi\frac{dA\,d\phi}{|\nabla\phi|}\\=\int\rho\frac{dQ}{d\phi}A_{\phi}\,d\phi,
\end{split}
\end{equation}
where we have used the relation between differential volume and area
\begin{equation}
    dV = \frac{dA\,d\phi}{|\nabla\phi|}.
\end{equation}
Equating Equation~\ref{eq:lhs_eval} to Equation~\ref{eq:rhs_eval}, taking the derivative with respect to $\phi$, and subsequently solving for $Q$ yields:
\begin{equation}
    v_s^2\frac{d}{d\phi} \big(\rho A_{\phi} \big) =\rho A_{\phi}\frac{dQ}{d\phi}\implies Q=v_s^2\ln(\rho A_{\phi}).
\end{equation}

Inserting this expression for $Q$ into Equation~\ref{eq:convective_momentum} and taking a dot product with $\vec{v}$ yields:
\begin{equation}
    \vec{v}\cdot\nabla\left[\frac{1}{2}v^2+\int\frac{dP}{\rho}+\phi+v_s^2\ln(\rho A_{\phi})\right]=0,
\end{equation}
The term inside the brackets is conserved along a streamline. Assuming isothermal ideal gas pressure in the chromosphere such that $P=c_T^2\rho$, we can write the one-dimensional constant as:
\begin{equation}\label{eq:mod_bern}
    \frac{1}{2}v^2+v_{\rm tot}^2\ln(\rho)+\phi+v_s^2\ln (A_{\phi})=\rm{const.},
\end{equation}
where
\begin{equation}
    v^2_{\rm tot} = v^2_s+c^2_T.
\end{equation}

At the stellar photosphere, $v \approx0$. Then, using the photospheric potential, $\phi_{\rm ph}$, and density, $\rho_{\rm ph}$, as the boundary conditions, Equation~\ref{eq:mod_bern} can be used to solve for the density at any potential. This yields:
\begin{equation}
    \rho(\phi,\,v_s)=\rho_{\rm ph}e^{-\frac{\phi-\phi_{\rm ph}}{v_{\rm tot}^2}-\frac{v^2}{2v_{\rm tot}^2}}\left[\frac{A_{\rm ph}}{A_{\phi}}\right]^{\frac{v_s^2}{v_{\rm tot}^2}}.
\end{equation}
Though we achieved a closed form solution for the density, the potential-dependent speed, $v\equiv v(\phi)$, has not yet been calculated. However, in analogy with \cite{Lubow75}, we expect the velocity near the L1 point to be $\vec{v} \simeq \vec{v}_s+\vec{c_T}$. We then have:
\begin{equation}
    \langle v^2\rangle=v_s^2+c_T^2+2\langle\vec{v_s}\cdot\vec{c_T}\rangle,
\end{equation}
where $\langle\rangle$ denotes a spatial average. For isotropic thermal motion of the gas particles, $\langle\vec{v_s}\cdot\vec{c_T}\rangle=0$, so $v^2 \simeq v_{\rm tot}^2$. The density near the L1 point thus takes the form:
\begin{equation}\label{eq:chrom_density_profile}
    \rho(\phi,\,v_s) \simeq \frac{1}{\sqrt{e}}\rho_{\rm ph}\,e^{-\frac{\phi-\phi_{\rm ph}}{v_{\rm tot}^2}}\left[\frac{A_{\rm ph}}{A_{\phi}}\right]^{\frac{v_s^2}{v_{\rm tot}^2}},
\end{equation}
A complete derivation for the density would involve solving the continuity and momentum equations consistently as done in \cite{Cehula23}. This would be more accurate and would not need a volume-averaged formalism as used here, but the resultant correction is expected to be of order unity, and, thus, a minor effect in our subsequent calculations. Thus, we use Equation~\ref{eq:chrom_density_profile} throughout our work. 

We now check two limiting cases:
\begin{itemize}
    \item Thermal-motion dominated atmosphere $(c_T \gg v_s)$: in this case, we obtain the density profile as 
    \begin{equation}\label{eq:rho_ritter}
        \rho(\phi) \simeq \frac{1}{\sqrt{e}}\rho_{\rm ph}e^{-\frac{\phi-\phi_{\rm ph}}{c_T^2}},
    \end{equation}
    which is the classical thermal pressure supported atmosphere used in the optically thin mass transfer calculation in \citetalias{Ritter88}.
    \item Shock-dominated chromosphere $(v_s\gg c_T)$: this yields a solely shock-supported atmosphere:
    \begin{equation}\label{eq:vs_gg_cT}
        \rho(\phi,\,v_s) \simeq \frac{1}{\sqrt{e}}\rho_{\rm ph}\frac{A_{\rm ph}}{A_{\phi}}e^{-\frac{\phi-\phi_{\rm ph}}{v_s^2}}.
    \end{equation}
    This is very similar to an isothermal atmosphere but with $c_T$ replaced by $v_s$, and with the additional geometric term $A_{\rm ph}/A_\phi$.
\end{itemize}

%We note that our treatment is expected to fail when the shock speed becomes large. This is because for high-speed shocks, assumption~2 will not be valid (for example, in the limiting case of very high velocity, a radial shock from the stellar surface will propagate ballistically without feeling the potential and escape the binary). Thus, this expression is expected to be valid only for the regime $v_s\lesssim\sqrt{2|\phi|}$. 

Assuming a shock speed distribution given by Equation~\ref{eq:vs_gauss}, one can then define a shock speed-averaged density profile:
\begin{equation}\label{eq:vs_avg_dens}
    \rho(\phi) = \int_0^{\infty}\rho(\phi,\,v_s)\,p(v_s)\,dv_s \, .
\end{equation}
We evaluate this for both single stars and Roche potentials below.

\subsubsection{Reviewing the single-star chromospheric profile}

\begin{figure}
    \centering
    \includegraphics[width=1\linewidth]{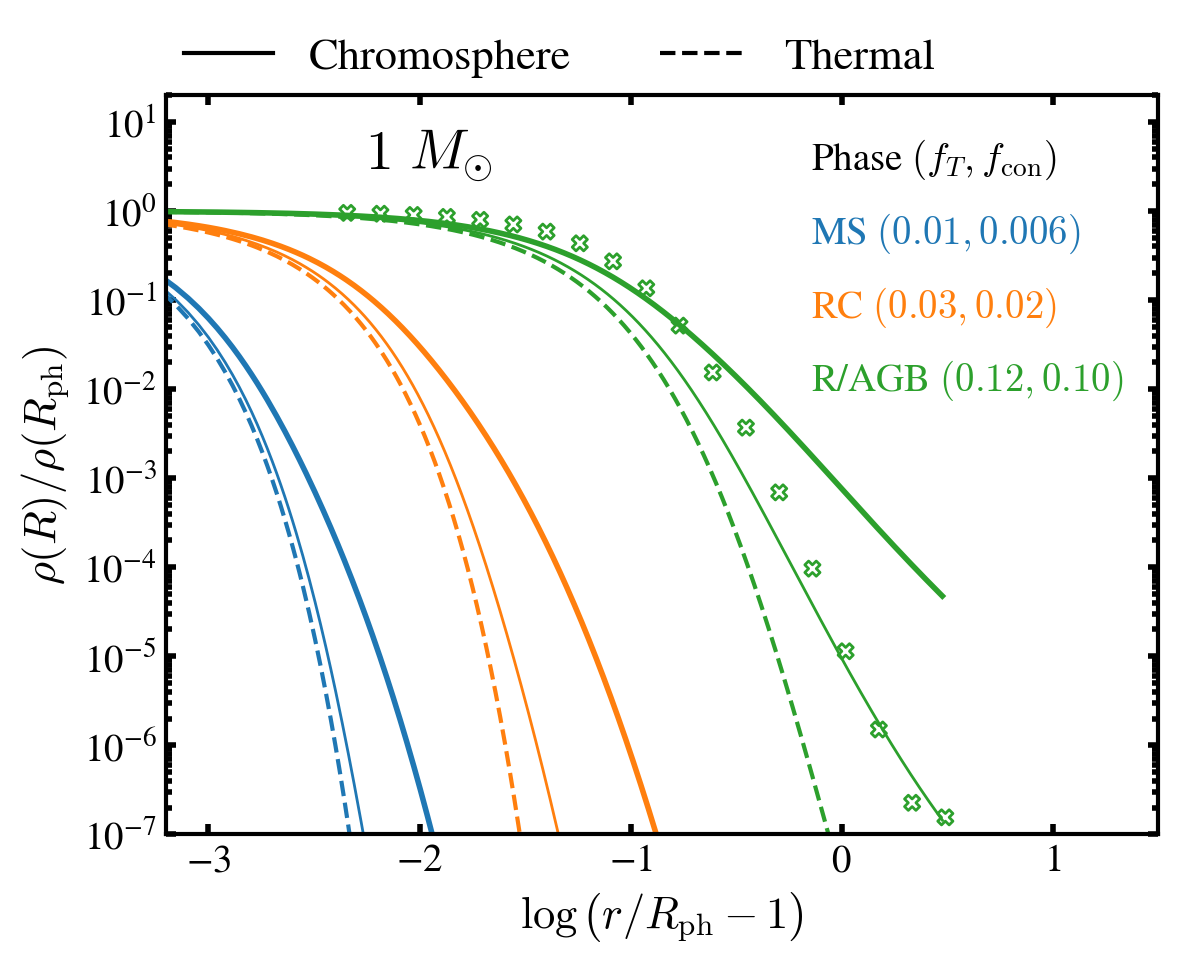}
    \caption{Chromospheric density profiles (solid lines, Equation~\ref{eq:density_profile_single_star}), normalized to its value at $R_{\rm ph}$, for three representative pairs of $f_T$ and $f_{\rm con}$ roughly corresponding to the main sequence (MS), red clump (RC) and tip of red giant branch or asymptotic giant branch (TRGB/AGB) phases of a $1~M_{\odot}$ MESA model. The thick and thin lines show the profiles with $f_{\rm con}$ (full chromosphere) and $0.5\times f_{\rm con}$ (reduced chromosphere), respectively. The dashed lines show the isothermal pressure-supported atmosphere profile for the same $f_T$ values. The green crosses are the density profiles from the hydrodynamical simulations of a solar-mass AGB star with a luminosity of $3911~L_{\odot}$ from \cite{Ahmad23} normalized to the photospheric density.}
    \label{fig:1msun_density_profile}
\end{figure}

A single star is simpler than a binary, so it is useful to first evaluate the chromosphere model for a single star potential. Assuming spherical equipotential surfaces and substituting $\phi(R)=-GM/R$ in Equation~\ref{eq:chrom_density_profile}, we get:
\begin{equation}\label{eq:density_profile_single_stars}
    \rho(R,\,v_s) \simeq \frac{1}{\sqrt{e}}\rho_{\rm ph}e^{-\frac{v_{\rm esc}^2}{2v_{\rm tot}^2}\left(1-\frac{R_{\rm ph}}{R}\right)}\left(\frac{R_{\rm ph}}{R}\right)^{\frac{2v_s^2}{v_{\rm tot}^2}}.
\end{equation}
In the limit of $v_s\gg c_T$, $v_{\rm tot}\rightarrow v_s$, this is essentially identical to equation~5 in \citetalias{Fuller24}.
\footnote{\citetalias{Fuller24} does not have the factor of $1/\sqrt{e}$, because they assumed hydrostatic equilibrium rather than computing the appropriate Bernoulli constant.}
%This is because, their approach intrinsically assumes the bulk velocity to be $v=v_{\rm tot}=v_s$ and constant everywhere. Using this in the Bernoulli parameter drops the $e^{-v^2/2v_s^2}$ term entirely from the density formula.
But Equation~\ref{eq:density_profile_single_stars} is a more general solution, incorporating both the shock and thermal support terms. 

The resultant shock speed-averaged profile (Equation~\ref{eq:vs_avg_dens}), using the parameters $f_T$ (Equation~\ref{eq:fT}) and $f_{\rm con}$ (Equation~\ref{eq:fcon}), can be written as:
\begin{equation}\label{eq:density_profile_single_star}
\begin{split}
    \rho(R) \simeq \sqrt{\frac{2}{\pi e}}\frac{\rho_{\rm ph}}{f_{\rm con}}\times\\\int_0^{\infty}\exp\frac{1}{2}\left[-\frac{1-\frac{R_{\rm ph}}{R}-4v'^2\ln\left(\frac{R_{\rm ph}}{R}\right)}{v'^2+f_T^2}-\frac{v'^2}{f_{\rm con}^2}\right]dv',
    \end{split}
\end{equation}

\begin{comment}
Averaging over the shock speed distribution, Equation~\ref{eq:vs_gauss}, the general form chromospheric density for a single star takes the form:
\begin{equation}
\begin{split}
    \rho(R)=\int_0^{\sqrt{2|\phi|}}\rho(\phi,~v_s)\,dv_s\\=\sqrt{\frac{2}{\pi e}}\frac{\rho_{\rm ph}}{v_{\rm con}}\int_0^{\sqrt{\frac{2GM}{R}}}e^{-\frac{v^2_{\rm esc}}{2v^2_{\rm tot}}\left(1-\frac{R_{\rm ph}}{R}\right)}\left(\frac{R_{\rm ph}}{R}\right)^{\frac{2v_s^2}{v^2_{\rm tot}}}\\\times e^{-\frac{v_s^2}{2v^2_{\rm con}}}\,dv_s.
    \end{split}
\end{equation}
The equation can be recast to a much cleaner and scale-free form using the parameters $f_T$ (Equation~\ref{eq:fT}) and $f_{\rm con}$ (Equation~\ref{eq:fcon}) to:
\end{comment}

\begin{comment}
\begin{figure}
    \centering
    \includegraphics[width=1\linewidth]{Figures/1msun2msun_mass_profile.png}
    \caption{Profile of overlying chromospheric mass as a function of radius for $1~M_{\odot}$ and $2~M_{\odot}$ AGB stars for different $f_{\rm con}$ parameters (the gray lines represent the thermal atmosphere). For both stars, we choose an epoch such that the photospheric density is $\rho_{\rm ph}=2\times10^{-9}~\rm g/cm^3$, which correspond to stellar radii of $210~R_{\odot}$ and $350~R_{\odot}$, respectively.}
    \label{fig:1msun2msun_mass_profile}
\end{figure}
\end{comment}

\noindent Though this is more accurate, it cannot be reduced to a closed form until a shock-dominated limit is assumed which sets $v_{\rm tot}=v_s$. This recovers equation~7 in \citetalias{Fuller24} which is numerically correct within a factor of a few.

%Note that this does not have a closed form solution, but presents a more accurate density profile of the chromosphere. However, the dominant contribution to the density profile comes from the shocks, i.e., the parameter $f_{\rm con}$. In almost all cases we investigate below, $f_{\rm con}\lesssim0.1$. Thus, $v_{\rm max}'=\infty$ is a reasonable approximation. Overall, the approximate closed form solution given by equation~7 in \citetalias{Fuller24} is correct within a factor of a few.

We present the density profile for a $1~M_{\odot}$ star in Figure~\ref{fig:1msun_density_profile}, for three representative stages of its life, characterized by different values of $f_T$ and $f_{\rm con}$ obtained from MESA models -- we use $\alpha_{\rm MLT}=2.0$ and, following \cite{Fuller24,Goldberg22}, we evaluate $v_{\rm con}$ where the optical depth is $\tau=P_{\rm rad}c/(Pv_{\rm con})$, where $P_{\rm rad}$ and $P$ are the radiation and gas pressures, respectively, and $c$ is the speed of light. For comparison, we also show the corresponding thermal pressure-supported atmospheric profiles. It is evident that shocks can support orders of magnitude denser chromospheres, extending up to several stellar radii during the giant phases.

However, comparison with the density profiles from the CO5BOLD simulations in \cite{Ahmad23} shows that the $v_{\rm con}$ values obtained from MESA might be over-predicting the densities. But, interestingly, using only half the $v_{\rm con}$ value yields better agreement. Henceforth, we use $v_{\rm con}/2$ for the `reduced chromosphere' case, against the `full chromosphere' case using the actual $v_{\rm con}$ values. This indicates that the $v_{\rm con}$ value to be used in our model is uncertain and needs appropriate calibrations through comparison to hydrodynamical simulations. We aim to undertake this exercise in future studies with additional simulations of AGB stars (for example, Glanz et al. in prep). For the purpose of this paper, we discuss the results with the reduced chromosphere whenever needed.

Note that the chromosphere also drives winds -- under the model of \citetalias{Fuller24}, the chromosphere
lifts off material to the dust formation radius, where the gas
condenses to dust and gives rise to the dust-driven winds. We call this the `Chromospheric Wind'. The prescription in the limit of $v_{\rm tot}=v_s$ for single stars has been derived in \citetalias{Fuller24}. But in binaries, the Roche potential will shape the wind formation and velocity. A part of the wind may also be channeled through the L1 nozzle. Appendix~\ref{app:coronal_wind} presents an implementation of models in which the standard wind prescriptions in MESA are replaced with the \citetalias{Fuller24} Chromospheric Wind. We defer calculation of the effect of the companion on the wind to future work.
  %We shall present this calculation in a future manuscript and keep this work focused on the region within the dust formation radius. We discuss a simple version of the Chromospheric wind and implement it in MESA in Appendix~\ref{app:coronal_wind}.}

%Figure~\ref{fig:1msun2msun_mass_profile} shows the mass contained in the chromosphere for $1~M_{\odot}$ and $2~M_{\odot}$ giants for three representative $f_{\rm con}$ values, compared to the thermal atmosphere ($f_{\rm con}=0$). The chromosphere can hold several times more mass compared to a thermal atmosphere. The total chromospheric mass increases with increasing stellar mass/radius (see also figure~3 in \citetalias{Fuller24}).

%for single stars, though a closed form solution for $v_s$-averaged density, like their equation~7, is no longer possible. 

\subsubsection{Chromosphere in a binary}\label{subsubsec:chrom_dens_binary}

We shall now evaluate Equation~\ref{eq:chrom_density_profile} for a binary potential. The gravitational potential $\phi$ in a binary system -- with donor mass $M_d$, accretor mass $M_a$, corresponding mass ratio $q=M_d/M_a$, and semimajor axis $a$ -- is given by the Roche potential (Appendix~\ref{app:quants}). For ease of calculation, henceforth, we primarily work with normalized (primed) quantities where the potentials, velocities, and lengths are normalized by $GM_d/a$, $\sqrt{GM_d/a}$, and $a$, respectively. 

The equipotential areas and the photospheric potential are primary inputs for Equation~\ref{eq:chrom_density_profile}. Unlike the spherical single star case, the Roche potential is not simple enough to derive closed form expressions for either of the quantities. Although numerically evaluating the area is the direct and accurate approach, it is computationally inefficient. Thus, we resort to an approximate approach of using volume equivalent spheres. For a given volume-equivalent radius, $R'^V$, equation~41 in \citealt{Jackson17} gives the corresponding potential:
\begin{equation}\label{eq:vol_eq_pot}
\begin{split}
    -\phi'^V(R'^V) = \frac{1}{q}+\frac{1}{2(q+q^2)}+\frac{1}{R'^V}+\frac{1+1/q}{3}\left(R'^V\right)^2\\+\frac{4}{45}\left[(1+1/q)^2+\frac{9}{q^2}+\frac{3}{q}(1+1/q)\right]\left(R'^V\right)^5,
    \end{split}
\end{equation}
where the superscript $V$ in $\phi'^V$ emphasizes the volume-equivalent approximation underlying the potential value.

For a given stellar photospheric radius, we use Equation~\ref{eq:vol_eq_pot} to obtain the photospheric potential $\phi'^V_{\rm ph}$. To obtain the equipotential areas above the photosphere, we solve Equation~\ref{eq:vol_eq_pot} for $R'^V_{\phi}$ at a given potential $\phi'^V$. We then simply use the surface area of this sphere as the corresponding equipotential surface area. Comparison with numerical evaluations shows that this agrees with the true area within $\lesssim$$10\%$ (Appendix~\ref{app:numerical_area}). With these approximations, Equation~\ref{eq:chrom_density_profile} can be recast to the following spherical-equivalent form in primed quantities:
\begin{equation}\label{eq:chrom_density_voleq}
    \rho^V(\phi',\,v'_s) \simeq \frac{1}{\sqrt{e}}\rho_{\rm ph}\,e^{-\frac{\phi'-\phi'^V_{\rm ph}}{v_{\rm tot}'^2}}\left[\frac{R'_{\rm ph}}{R'^V_{\phi}}\right]^{\frac{2v_s'^2}{v_{\rm tot}'^2}},
\end{equation}
which we use in our subsequent calculations. We then integrate over shock speeds (as in Equation \ref{eq:vs_avg_dens}) to obtain the total density.

\begin{figure}
    \centering
    \includegraphics[width=1\linewidth]{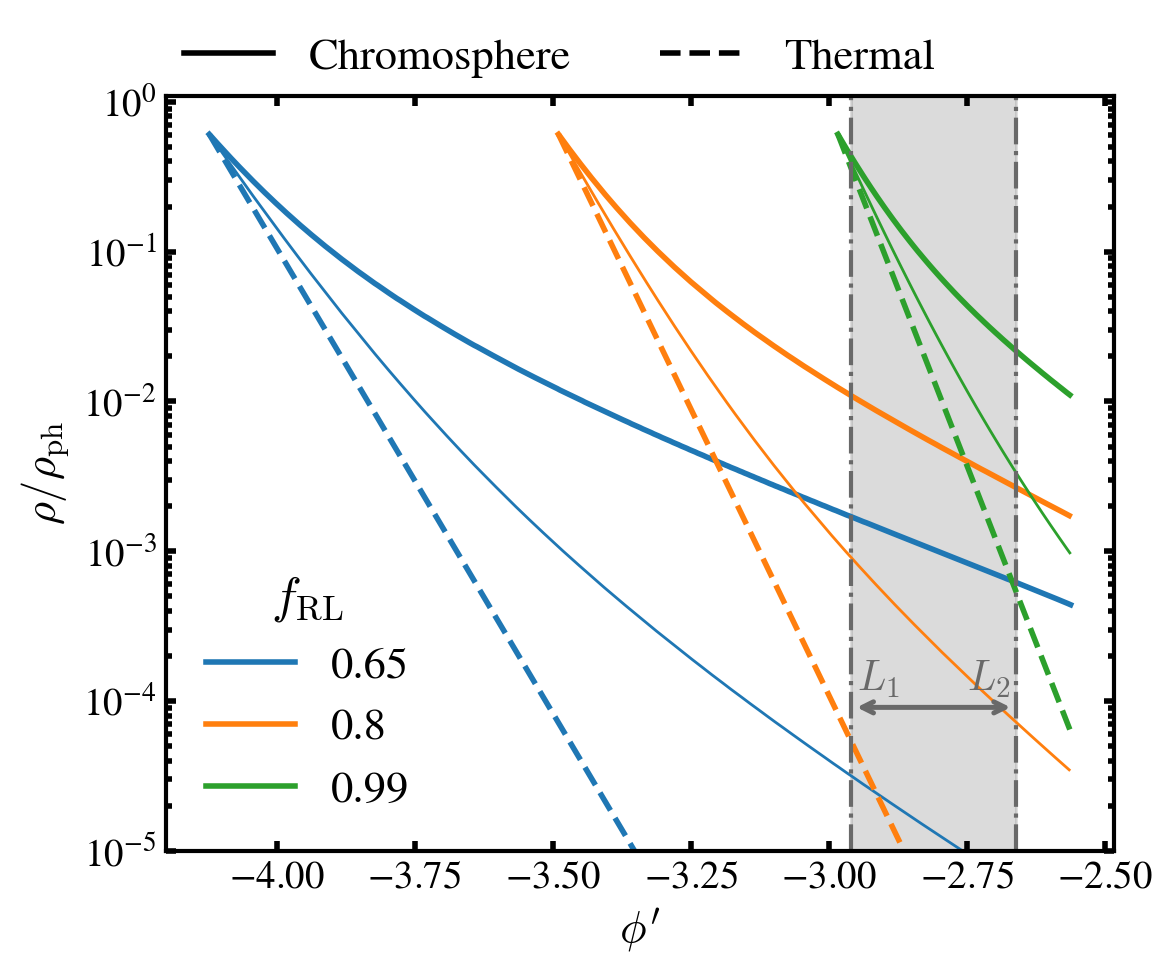}
    \caption{Chromospheric density profile (solid lines) in a binary potential of donor mass $M_d=1\,M_{\odot}$, accretor mass $M_a=0.5\,M_{\odot}$, and orbital period of $P_{\rm orb}=10^3$~days as a function of the normalized potential for three values of Roche fill factors, $f_{\rm RL}$. The thick and thin lines are for the full and reduced chromospheres, respectively. We also present the isothermal pressure-supported atmospheric profiles (dashed lines) for comparison. We assume representative values of $f_T=0.1$ with $f_T=f_{\rm con}=0.1$. The region between the L1 and L2 potentials -- primarily responsible for mass transfer -- is shaded. The chromosphere allows for substantially denser material to reach the L1 nozzle.}
    \label{fig:rho_chrom_binary}
\end{figure}

Figure~\ref{fig:rho_chrom_binary} shows the shock speed-averaged chromospheric density profile, calculated using Equations~\ref{eq:vs_avg_dens} and \ref{eq:chrom_density_voleq}, for a representative binary. We present the results for three values of Roche fill factors:
\begin{equation}
    f_{\rm RL}=R_{\rm ph}/R_{\rm L1},
\end{equation}
where $R_{\rm L1}$ denotes the volume-equivalent Roche lobe radius of the L1 equipotential \citep[][see Appendix~\ref{app:quants}]{Eggleton83}. Larger $f_{\rm RL}$ signifies that the stellar photosphere is closer in potential to the L1 point. We find that the chromosphere allows orders of magnitude denser material to reach the Lagrange potentials compared to a thin thermal atmosphere, thus enhancing binary interaction substantially.

\subsection{Mass Transfer Rate}\label{subsec:chrom_mdot}

We now use the density profile derived in the previous sections to calculate the mass transfer rate using Equation~\ref{eq:mdot_basic}. An important aspect is the potential window over which the mass transfer takes place, which sets the limit of the integral. This is uncertain, but we reason that the companion accretes from the matter that remains bound to the stellar potentials. Starting from the L1 potential, $\phi_{\rm L1}$, the maximum potential which meets this criterion is the minimum of the L2 and L3 potentials -- $\phi_{\rm max}=\min(\phi_{L2},~\phi_{L3})$ -- beyond which the matter is ejected. We adopt this range of potentials as our channel for mass transfer. Additionally, as in Section~\ref{subsec:chrom_dens_profile}, the speed $v = v_{\rm tot} = \sqrt{ v_s^2 + c_T^2}$ is the characteristic flow speed through the L1 nozzle.

Using Equation~\ref{eq:chrom_density_voleq} in Equation~\ref{eq:mdot_basic} and convolving with the shock-speed distribution (Equation~\ref{eq:vs_gauss}), the mass transfer rate then becomes in terms of the primed quantities:
\begin{equation}\label{eq:mdot_chrom_condensed}
\begin{split}
    \dot{M} = \sqrt{GM_da^3}\\\times\int_{\phi'_{L1}}^{\phi'_{\rm max}}\int_0^{\infty}\rho^V(\phi',\,v'_s)\,v'_{\rm tot}\,\frac{dA'_N}{d\phi'}\,p(v'_s)\,dv'_s\,d\phi',
\end{split}
\end{equation}
where $A'_N=A_N/a^2$. Note that Equation~\ref{eq:chrom_density_voleq} is only valid for $\phi>\phi^V_{\rm ph}$; thus, Equation~\ref{eq:mdot_chrom_condensed} will result in erroneous $\dot{M}$ when the star (over)fills its Roche lobe. To bridge this gap, we define the ``saturated'' chromospheric accretion rate by setting $\rho^V(\phi<\phi_{\rm ph})=\rho^V(\phi_{\rm ph})$. For the nozzle area, we use equation~8 in \cite{Marchant21}:
\begin{equation}\label{eq:dAN_by_dphi}
\begin{split}
    \frac{dA_N'}{d\phi'}=\frac{\pi}{\sqrt{C_1'C_2'}}\left(1-\frac{C_3'(3C_1'^2+2C_1'C_2'+3C_2'^2)}{4C_1'^2C_2'^2}\Delta\phi'\right)
    \end{split}
\end{equation}
where $\Delta\phi'=\phi'-\phi'_{L1}$ and the constants are provided in Appendix~\ref{app:quants}. This agrees with the numerical nozzle area up to $\phi'_{\rm max}$ within $<$$5\%$ (Appendix~\ref{app:numerical_area}). The double integration in Equation~\ref{eq:mdot_chrom_condensed}, however, often proves to be numerically expensive. We thus use a reduced single integral correct within a few percent, the details of which are provided in Appendix~\ref{app:reduced_chrom_mdot}.

\begin{figure}[t]
    \centering
    \includegraphics[width=1\linewidth]{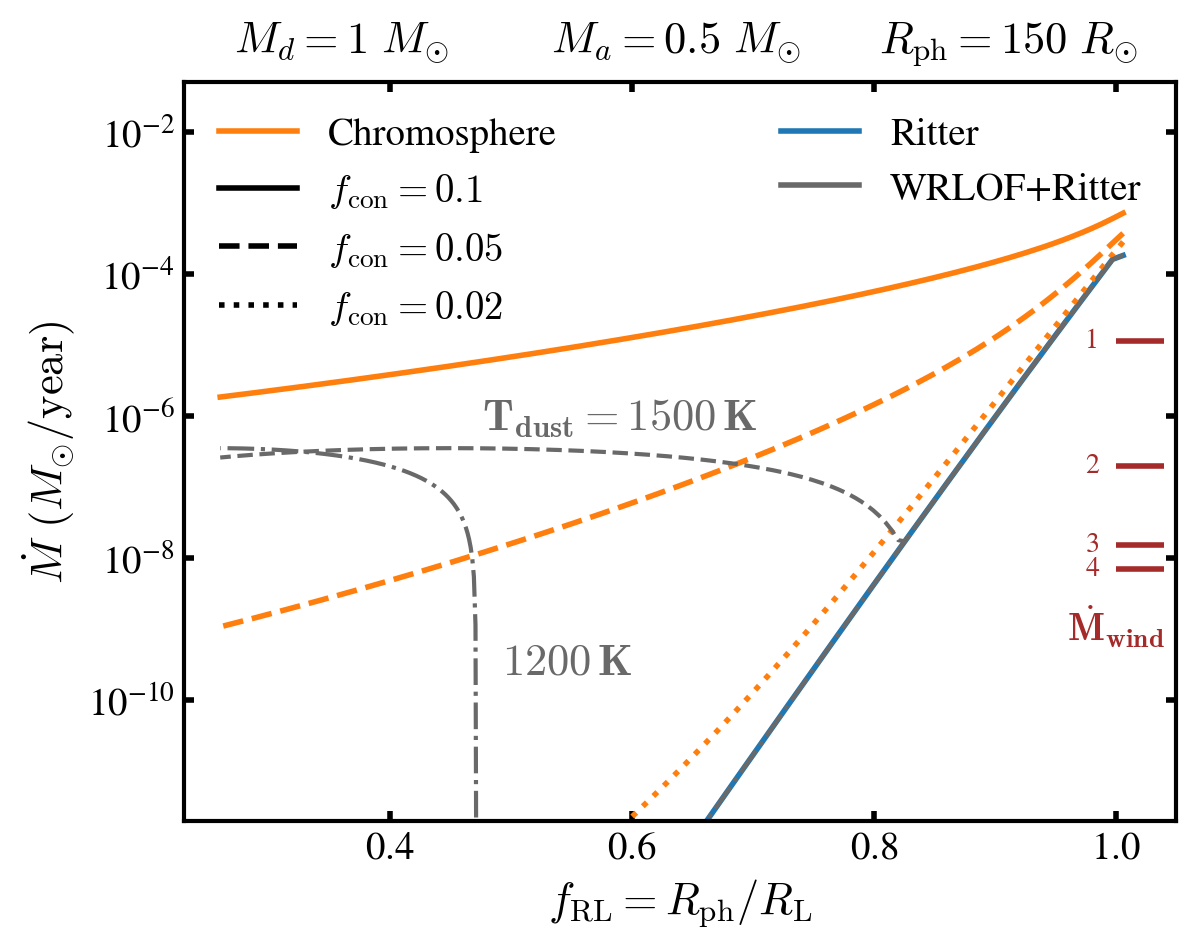}
    \caption{The chromospheric $\dot{M}$ compared to the classical \citetalias{Ritter88} prescription for a red giant with a mass of $M_d = 1 \, M_\odot$ and fixed $R_{\rm ph} = 150 \, R_\odot$, $T_{\rm ph}=3300$~K, and $L\approx2400~L_{\odot}$ as a function of the Roche fill factor. Such a giant is expected to have $f_{\rm con}\approx0.08$; nevertheless we show the results for different $f_{\rm con}$ values for comparison. In the Roche under-filling regime, even moderate chromospheric rates are orders of magnitude higher than the \citetalias{Ritter88} prescription. Our prescription converges to the \citetalias{Ritter88} rate when $f_{\rm con}\rightarrow0$. The grey lines shows the mass transfer rate from the Wind Roche Lobe Overflow (WRLOF) model, $\dot{M}_{\rm WRLOF}=\beta_w\dot{M}_{\rm wind}$ (see Section \ref{sec:wrlof}) with a fiducial $\dot{M}_{\rm wind}=10^{-6}~M_{\odot}/{\rm yr}$ and two cases of assumed dust formation temperatures, $T_{\rm dust}$. The left end of the plot is approximately the dust formation radius and, thus, the validity limit for the chromospheric model. For comparison, we mark the $\dot{M}_{\rm wind}$ from the four wind models in Appendix~\ref{app:coronal_wind} with brown dashes at the donor's luminosity. They are, in the numerical order, Full Chromospheric, Bloecker \citep{Blocker95}, Reimers \citep{Reimers75}, and Reduced Chromospheric winds.}
    %added to the \citetalias{Ritter88} rate. It agrees with the chromospheric rates at low $f_L$, but the drop in $\dot{M}$ for intermediate $f_L$ appears unphysical.
    \label{fig:mdot_1msun_q2_fixedRd}
\end{figure}

Figure~\ref{fig:mdot_1msun_q2_fixedRd} demonstrates and contrasts the chromospheric mass transfer rate (henceforth, $\dot{M}_c$) with the optically thin \citetalias{Ritter88} prescription (henceforth, $\dot{M}_R$). $\dot{M}_c$ is exponentially dependent on the convective velocity parameter $f_{\rm con}$. But, for moderate values of $f_{\rm con}$ appropriate for stars high on the RGB or AGB, $\dot{M}_c$ can be orders of magnitude higher than $\dot{M}_R$, particularly for low Roche fill factors. This can substantially affect binary evolution as wide systems which would have escaped any interaction in the classical framework may instead undergo significant episodes of mass transfer when the extended chromosphere is taken into account.

\section{Chromospheric mass transfer in MESA}\label{sec:mesa_chrom}

\begin{figure*}[t]
    \centering
    \includegraphics[width=1\linewidth]{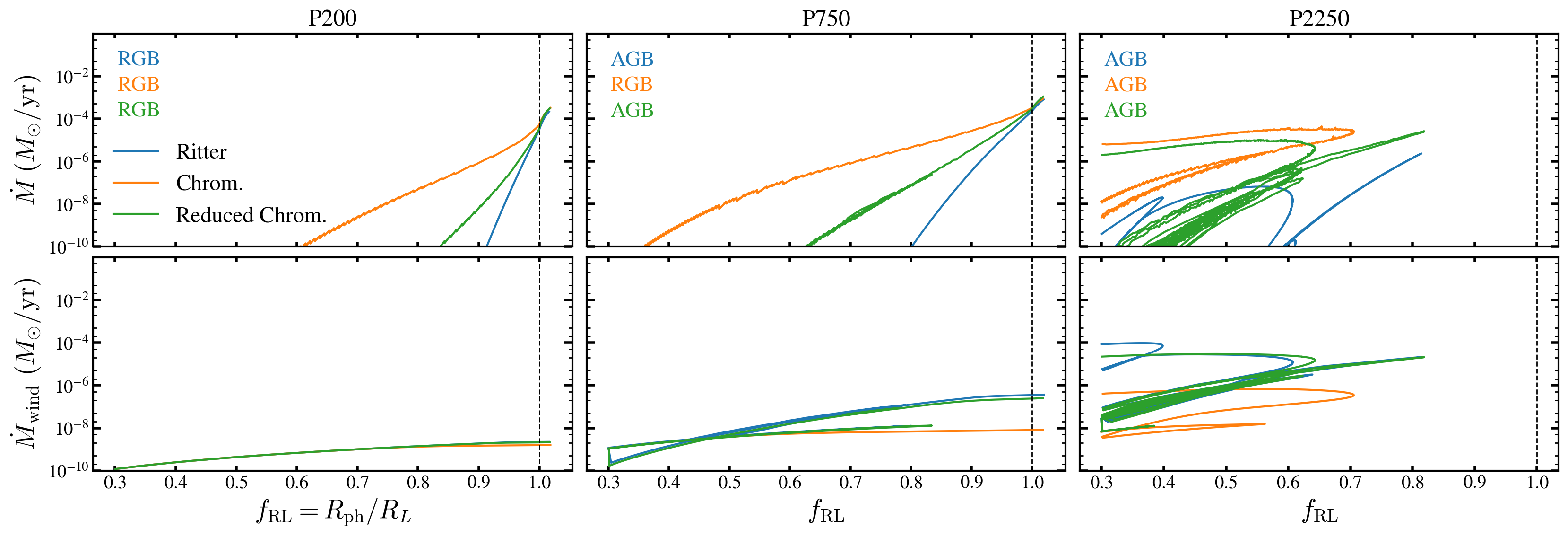}
    \caption{Mass transfer rate (top panel) as a function of the Roche filling factor for three MESA runs using both optically thin accretion prescriptions -- \cite[][blue]{Ritter88}, our full chromosphere model (this work, orange), and our reduced chromosphere model (this work, green). We also indicate in correspondingly colored text whether the mass transfer and envelope stripping occurred during the RGB or AGB phase. The vertical dashed line marks the onset of the optically thick regime (we truncate the plot at $f_{\rm RL}=1.02$ to focus on the optically thin $f_{\rm RL}<1$ regime). The chromospheric prescription yields orders of magnitude larger accretion rates in the optically thin regime. In the P2250 case, the mass transfer is driven by the thermal pulses, resulting in the multiple ``wiggles'' in the plot. For comparison, we also show the corresponding wind mass loss rates in the bottom panel following the Reimers/Bloecker prescriptions. See Appendix~\ref{app:coronal_wind} for chromospheric winds.}
    \label{fig:quants_v_rdbyrl}
\end{figure*}

\begin{figure*}[t]
    \centering
    \includegraphics[width=1\linewidth]{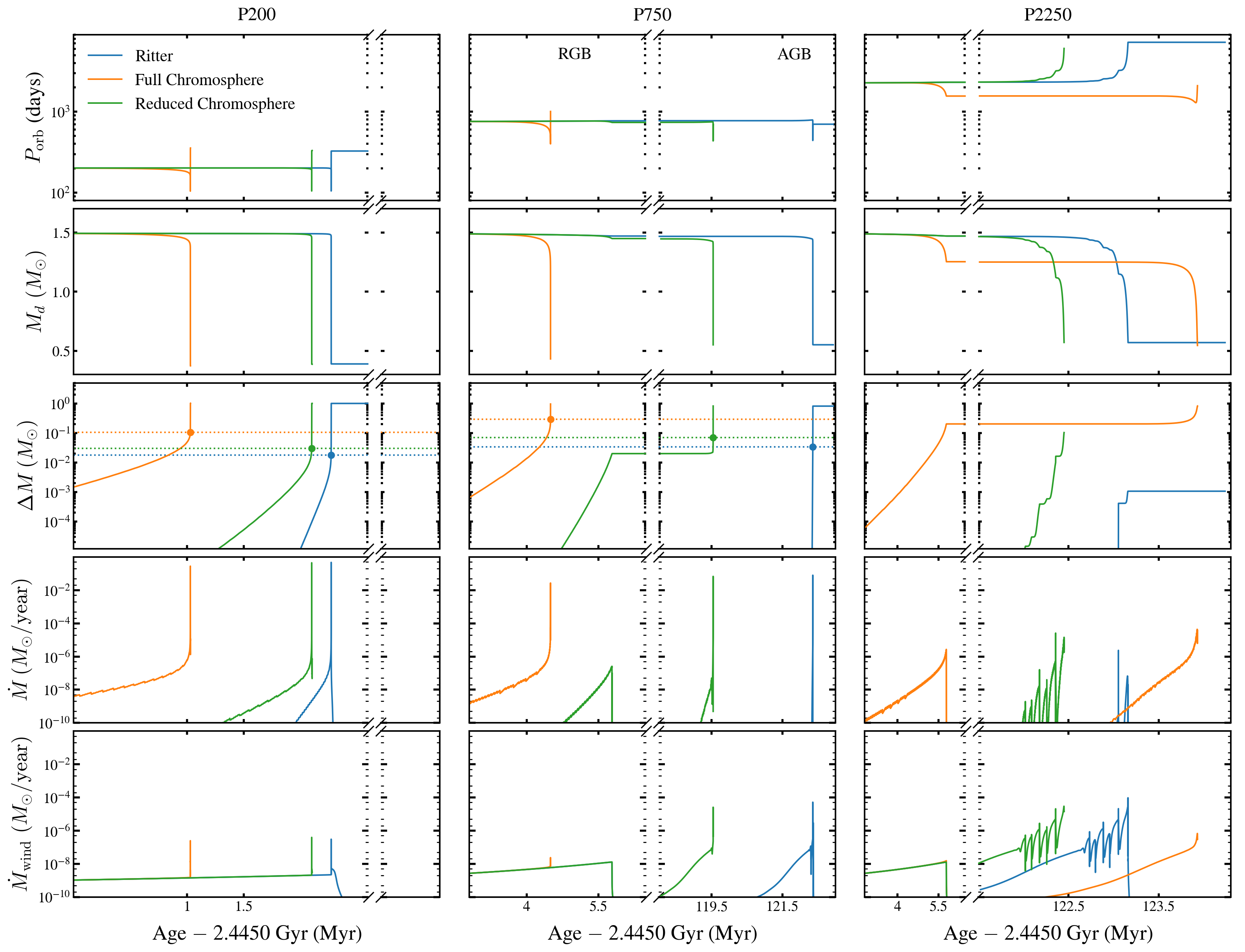}
    \caption{Stellar and binary parameters as a function of age for our MESA runs. From top to bottom, we show the evolution of binary period, $P_{\rm orb}$; the donor mass $M_d$; the total mass transferred, $\Delta M$; and the L1 mass transfer rate $\dot{M}$, and the wind mass loss rate $\dot{M}_{\rm wind}$. For better visualization, we zoom in on the upper RGB and AGB phases across a gap in time. In the third row, we also mark the epoch where $f_{\rm RL}=1$ with a circle. For the chromospheric runs, MESA crashes as the donor envelope gets stripped off quickly through high mass transfer rates. Overall, the extended chromosphere allows for enhanced interaction, especially for wide orbits. }
    \label{fig:quants_v_age}
\end{figure*}

In this section, we demonstrate the possible effects of chromospheric mass transfer on binary evolution using the stellar evolution code MESA \citep{Paxton11,Paxton15,Paxton19}. We use the latest MESA version of \texttt{r24.03.1}. For the purpose of this work, we consider the accretor to be a point mass by setting \texttt{evolve\_both\_stars=.false.} -- the evolution of the accretor will be studied in future work. We implement our prescription for $\dot{M}_c$ by setting \texttt{use\_other\_rlo\_mdot = .true.}, which replaces the optically thin \citetalias{Ritter88} prescription with the chromospheric accretion when $R_{\rm ph}<R_L$. When $R_{\rm ph}>R_L$, we continue to use the optically thick \cite{Kolb90} prescription, in combination with the saturated chromospheric mass transfer rate. We use fiducial donor and accretor masses of $1.5~M_{\odot}$ and $0.6~M_{\odot}$, respectively. We run three models with initial orbital periods of $200$~days, $750$~days and $2250$~days, which we call P200, P750, and P2250, respectively. We calculate models for both the original and the reduced ($0.5\times$) $f_{\rm con}$ values.

For comparison, we also run each of the models with the classical \citetalias{Ritter88} prescription for optically thin mass transfer. However, we realized that its current implementation in MESA (under the setting of \texttt{mdot\_scheme=`Kolb'/`Ritter'}) is incorrect. MESA currently implements equation~9 in \citetalias{Ritter88} which uses an approximate radial scale-height form of the density profile, which is only accurate for $R_{\rm ph}\approx R_{L1}$. However, the more accurate prescription is given by equation~A1 in \cite{Kolb90}, which uses the more generally valid density profile given by Equation~\ref{eq:rho_ritter}. The two expressions are identical when $R_{\rm ph} = R_{L1}$, but the current MESA implementation can underestimate $\dot{M}$ by orders of magnitude at low Roche fill factors (see Appendix~\ref{app:mesa_bonk}). Thus, we correct it in this work under the \texttt{other\_rlo\_mdot} hook and propose that this corrected implementation be made the default in MESA. As with the chromospheric mass transfer case, here too we use Equations~\ref{eq:chrom_density_voleq} and \ref{eq:eggleton} to compute the relevant potentials.

We use a solar metallicity of $Z=0.014$ for both the stars. We use the \cite{Ledoux47} criterion to determine convective instability, use a mixing length \citep{Cox68} parameter of $\alpha=2.0$, and we include semiconvection with $\alpha_{\rm sc}=0.1$ \citep{Choi16}. We also model winds following \cite{Choi16}: \cite{Reimers75} wind for the RGB phase with a scaling factor of $\eta_R=0.1$ and \cite{Blocker95} wind for the AGB phase with $\eta_B=0.2$. We keep here the conventional wind choices to concentrate on the effects of the change in the mass transfer prescription. The runs with Chromospheric winds are presented in Appendix~\ref{app:coronal_wind}. Additionally, we assume non-conservative mass transfer, with only $25\%$ of the transferred mass being accreted (i.e., $\beta = 0.25$; \citealt{Soberman97}); the rest is lost from the vicinity of the accretor.\footnote{Note that MESA defines $\beta$ as the fraction of the mass lost instead, so the \texttt{inlists} should be read accordingly.} We also include convective overshoot with \texttt{f=0.129} and \texttt{f0=0.0129} for the core, and the \texttt{exponential} scheme for the shell with \texttt{f=0.0174} and \texttt{f0=0.00174}. We use \texttt{energy\_eqn\_option = 'dedt'}.

Figure~\ref{fig:quants_v_rdbyrl} shows the mass transfer rate as a function of the Roche filling factor, $f_{\rm RL}$. This is the equivalent of Figure~\ref{fig:mdot_1msun_q2_fixedRd}, but with $f_{\rm con}$ computed self-consistently from the MESA models. We focus on the optically thin regime ($f_{\rm RL}<1$) and compare the classical and the new prescriptions. The extended chromosphere leads to substantially enhanced interaction, with mass transfer rates that can be orders of magnitude larger than the Ritter prescription when $f_{\rm RL} \lesssim 0.9$. For a given $f_{\rm RL}$, the chromospheric $\dot{M}$ increases with increasing initial orbital period. This is because longer periods require more evolved donors for the same $f_{\rm RL}$, and thus higher $f_{\rm con}$ values resulting from both higher $v_{\rm con}$ and lower $v_{\rm esc}$. 

Figure~\ref{fig:quants_v_age} then shows the evolution of the key stellar and binary parameters with age. Overall, the chromosphere allows for the onset of interaction at a smaller donor radius, leading to a more gradual ramp-up of the transferred mass until $f_{\rm RL}=1$ and the optically thick mass transfer starts. %Among other quantities, we also show the mass of carbon transferred to compare with carbon-enhanced stars which are thought to be post-AGB mass transfer products.
We briefly discuss the three initial period cases individually. 

\paragraph{P2250} Very little interaction takes place in the \citetalias{Ritter88} case, with a total transferred mass of $\approx$$10^{-3}~M_{\odot}$. Most of the donor envelope is ejected via wind, widening the period to $7227$~days with a resultant WD mass of $0.57~M_{\odot}$. The chromosphere leads to substantially more transferred mass. The full and reduced chromosphere cases lead to $\Delta M\approx1~M_{\odot}$ and $\approx$$2\times10^{-1}~M_{\odot}$, respectively -- orders of magnitude more than the \citetalias{Ritter88} case. In fact, in the full chromosphere case, the whole of the envelope gets stripped through mass transfer even solely through the optically thin phase. The enhanced mass transfer leads to shorter final orbits of $2100$~days and $6067$~days, respectively. The final WDs are $0.55~M_{\odot}$ and $0.57~M_{\odot}$, respectively.

\paragraph{P750} This is a notable case as the interaction proceeds differently for all three cases. In the \citetalias{Ritter88} case, interaction occurs only during the AGB with $\approx$$3\times10^{-2}~M_{\odot}$ transferred in the optically thin regime, followed by a substantial optically thick interaction resulting in a total transferred mass of $\approx$$1~M_{\odot}$. In the reduced chromosphere case, $\approx$$2\times10^{-2}~M_{\odot}$ is already transferred during the RGB phase, followed by a more prominent AGB interaction that transfers a total of $\approx$$8\times10^{-2}~M_{\odot}$ through the chromosphere in the optically thin regime, followed by an optically thick phase that transfers a total mass similar to that in the \citetalias{Ritter88} case. Both cases form a $\approx$$0.55~M_{\odot}$ CO-core WD at the end. The full chromosphere case, however, unfolds entirely differently as the donor gets completely stripped on the RGB itself, forming a $0.43~M_{\odot}$ He WD, with the optically thin chromosphere transferring a large $\approx$$0.3~M_{\odot}$ worth of mass, followed by an optically thick phase. This shows that the chromosphere can not only alter the amount of mass transferred, but also substantially influence the outcome of the interaction. The final period for the \citetalias{Ritter88} and reduced chromosphere cases is $\sim$$700$~days, whereas the RGB mass transfer with the full chromosphere leads to a $\sim$$1000$~day period. But the peak optically thick $\dot{M}$ achieves extremely high values of $>$$10^{-1.5}~M_{\odot}\rm/year$ in all cases which might trigger CEE resulting in final periods of $\approx$$1-3$~days (see Appendix~\ref{app:post_cee} for the details of the post-CEE period calculation).

%The RGB mass reduction leads to a delay in the AGB evolution, thus the AGB mass transfer occurs at a later time compared to the \citetalias{Ritter88} case. 

%This is a notable case as with the Ritter prescription mass transfer occurs in the AGB phase, whereas the chromospheric mass transfer leads to an envelope stripping at the RGB phase itself. Thus, the latter forms a He WD of mass $0.44~M_{\odot}$ rather than a $0.55~M_{\odot}$ CO-core WD in the Ritter case. This suggests that chromospheric mass transfer can result in early and/or enhanced stripping of the donor. We find that a large amount of mass of $\approx$$0.3~M_{\odot}$ is transferred during the optically thin phase from the chromosphere, which is a significant fraction of the donor envelope and about an order of magnitude larger than in the Ritter model. In both cases, the mass transfer eventually enters the optically thick regime where the peak $\dot{M}$ reaches very high values of $>$$10^{-1.5}~M_{\odot}\rm/year$. At such high rates, the stability of mass transfer is uncertain and the system may undergo a CEE. With a fiducial assumption that CEE sets in when $\dot{M}$ attains $10^{-2}~M_{\odot}\rm/year$, the period is $\approx$$1-3$~days (see Appendix~\ref{app:post_cee} for the details of the post-CEE period calculation). But, assuming stability and following the MESA model, the final orbital periods are similar at $703$~days and $990$~days, respectively.

\paragraph{P200} In this case, the outcome for all cases is broadly similar. The initial orbital period is short enough that interaction occurs during the RGB, with the final WD mass being $\approx$$0.38~M_{\odot}$. In all cases the binary would almost certainly enter a CEE resulting in post-CEE periods of $<$$1$~day -- rendering the MESA outcomes unreliable. The chromosphere, however, still allows for enhanced optically thin mass transfer with $\approx$$0.1~M_{\odot}$ and $\approx$$0.03~M_{\odot}$ in the full and reduced chromosphere cases, compared to $\approx$$0.015~M_{\odot}$ in the \citetalias{Ritter88} case. Additionally, denser chromospheres trigger earlier onsets of mass transfer.

%As with P750, the chromosphere transfers $\approx$$0.1~M_{\odot}$ mass in the optically thin regime, about $5$$\times$ than in the Ritter model. In the optically thick regime, the mass transfer rates attain extremely high values of $\gtrsim$$10^{-0.5}~M_{\odot}\rm/year$. Thus, the system almost certainly undergoes a CEE, rendering the MESA outcomes unreliable. We estimate a post-CEE period of $<$$1$~day.

\section{Discussion}\label{sec:discussion}

\subsection{Comparing with Wind Roche Lobe Overflow}
\label{sec:wrlof}

%At this stage, we point out that an existing theory for mass transfer beyond the classical channel is the Wind Roche Lobe Overflow \citep[WRLOF][]{Mohamed07} model. The model is based on the idea that the wind from the star fills the Roche lobe and gets channeled through the L1 nozzle. Although conceptually similar to this work and broadly successful in explaining observations \citep{Abate2013,Sun24}, there are certain limitations with the current implementation of WRLOF: 1) The prescription for $\dot{M}$ is empirically calibrated against a few smooth particle hydrodynamics (SPH) simulations \citep{Abate2013}, which may not be universally applicable. 2) Similar to the radiation-driven wind beyond the dust formation radius, the matter within is also treated as accelerating wind particles in the simulations. However, we expect the latter to be in gas phase and obey hydro(static)dynamic equations (thus forming a chromosphere) rather than particle motion. 

Our chromospheric mass transfer model is not the first model predicting significant interaction at low Roche filling factors. An existing model is the Wind Roche Lobe Overflow \citep[WRLOF][]{Mohamed07} model. Similar to our model, the WRLOF model is based on the idea that a fraction of the donor's wind gets channeled through the L1 nozzle -- particularly when the wind velocity, $v_w$, is smaller than the orbital velocity, $v_{\rm orb}$, i.e., when the wind is bound to the stellar potentials. The basis of this model is Smoothed Particle Hydrodynamics (SPH) simulations which model the wind as accelerating mass particles. Even when the star is under-filling the Roche lobe, the wind particles reach the L1 potential and facilitate L1-channeled mass transfer, thus the name. \cite{Mohamed07} found that the resultant $\dot{M}$ is orders of magnitude higher than that given by the standard Bondi-Hoyle-Lyttleton wind accretion prescription. Following their work, \cite{Abate2013} used the data points from the SPH simulations to derive an empirical accretion efficiency, $\beta_w$ (equation~9 in their paper), such that $\dot{M}_{\rm WRLOF}=\beta_w\dot{M}_{\rm wind}$, where $\dot{M}_{\rm wind}$ is the wind mass loss rate, and $\beta_w$ is a strong function of the assumed dust formation temperature, $T_{\rm dust}$. This prescription has been widely used in explaining systems like the formation of carbon enhanced stars \citep{Abate2013} or blue stragglers/lurkers \citep{Sun24}. 

Although WRLOF is conceptually similar to our model, we believe that there are certain aspects of the current implementation of WRLOF which are not well explained. In particular, the prescription suggests that the accretion efficiency decreases at large Roche filling factors, which seems unphysical. We demonstrate this in Figure~\ref{fig:mdot_1msun_q2_fixedRd} where we show a representative $\dot{M}_{\rm WRLOF}$ and compare it to \citetalias{Ritter88} and chromospheric mass transfer rates. Although at low $f_{\rm RL}$ the $\dot{M}$ from WRLOF is similar to our models, the WRLOF rate unexpectedly decreases with increasing $f_{\rm RL}$, until the Ritter prescription takes over. This is contrary to the general expectation that the accretion rate should monotonically increase with increasing $f_{\rm RL}$. Additionally, the model does not distinguish between the total mass transferred through the nozzle and the amount actually accreted onto the companion. For the purpose of our comparison, we simply assumed that they are the same (at least within a factor of a few).

Unfortunately, owing to the lack of public availability of the details of the simulations, particularly the PhD thesis of S. Mohamed that is cited in \cite{Abate2013}, we could not perform a detailed comparison. Without a clear explanation for the drop in mass transfer rate at large $f_{\rm RL}$, we believe those results may be affected by inaccuracies of the SPH simulations. Additional hydrodynamic simulations can help clarify this issue and resolve disagreements between our model and the WRLOF model. We note here, however, that while our calculation yields the mass transfer rate, the WRLOF prescription gives the actual accretion rate onto the accretor. But for this alone to explain the difference from our calculations, the accretion efficiency has to drop to very low fractions, which is not expected in the regime of such low mass transfer rates.

\subsection{Understanding post-mass transfer binaries}

\begin{figure}
    \centering
    \includegraphics[width=1\linewidth]{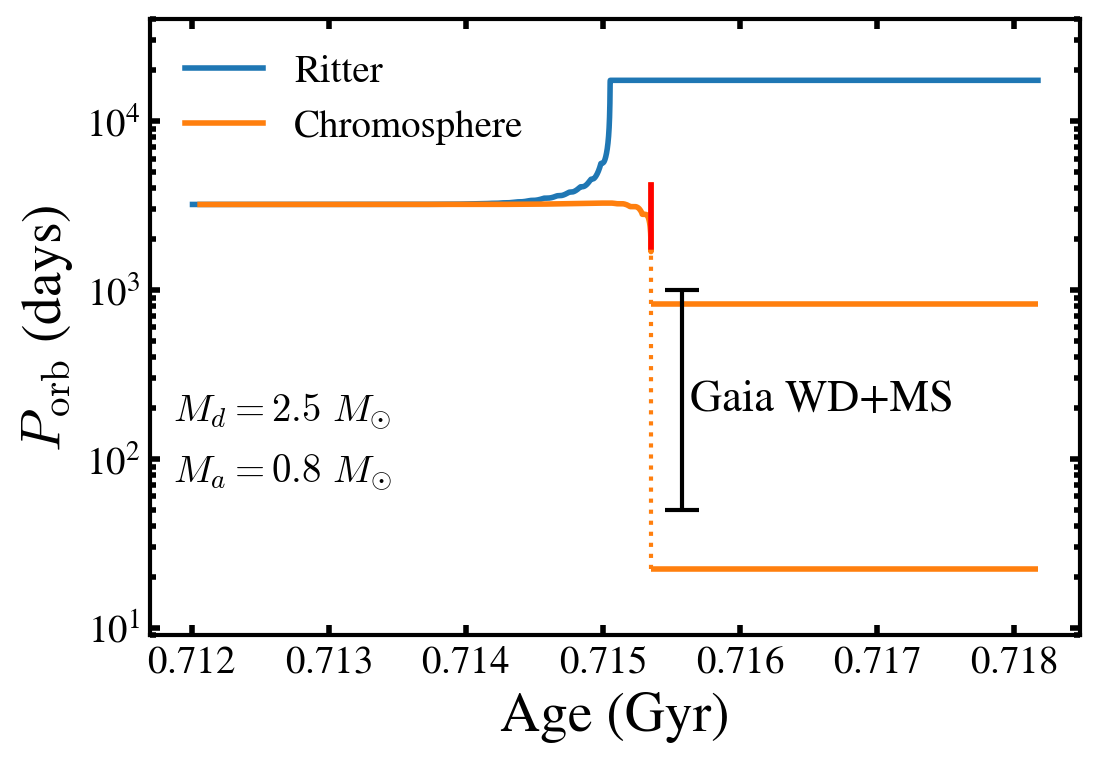}
    \caption{Orbital period evolution of a $2.5~M_{\odot}$ donor $+$ $0.8~M_{\odot}$ accretor with an initial period of $3200$~days and $\beta = 0.75$ for both the \citetalias{Ritter88} (blue) and full chromosphere (orange) prescriptions. The solid lines show the MESA outputs. The assumed epoch of CEE is when $\dot{M}$ reaches an arbitrarily high value of $10^{-1.5}~M_{\odot}/\rm year$. The red line marks the portion of the MESA output after this epoch. The orange horizontal lines at $P_{\rm orb} = 825.1$~days and $P_{\rm orb} = 21.4$~days mark the expected post-CEE period for the case of $\alpha_{\rm rec} = 1.0$ and  $\alpha_{\rm rec} = 0$, respectively. The post-CEE WD and the accretor masses are $0.59~M_{\odot}$ and $1.05~M_{\odot}$, respectively. The black bar shows the range of \textit{Gaia} WD-MS periods. This is an example of how chromospheric mass transfer can help produce post-CEE binaries similar to the \textit{Gaia} population (\citealt{Shahaf23,Yamaguchi24b}). }
    \label{fig:gaia_wdms_from_cee}
\end{figure}

%The enhanced interaction facilitated by the chromosphere may significantly impact our understanding of mass transfer stability and, consequently, outcomes of binary interaction. To understand why, we first provide a qualitative review of the evolution of the binary orbit upon mass transfer, particularly when the donor is more massive than the accretor (i.e., $q>1$). As mass transfer begins, the orbit shrinks initially, further enhancing $\dot{M}$. However, this only continues until the mass ratio reverses, owing to the donor (acretor) losing (gaining) mass. Post this, the orbit widens and $\dot{M}$ decreases, thus stabilizing the subsequent evolution.

%The results from Section~\ref{sec:mesa_chrom} shows that the enhanced interaction facilitated by chromosphere can substantially change the outcomes of a binary -- the amount of mass transferred as well as the final masses and periods. This can significantly impact our understanding of several classes of post-mass transfer systems. Though a population-level analysis is beyond the scope of this paper, we list a few examples where chromospheric mass transfer is expected to play significant role.

The discovery of populations of AU-scale ($P_{\rm orb}\sim10^2-10^3$~days) MS+WD/NS binaries in Gaia DR3 \citep{Yamaguchi24b,Yamaguchi25,Elbadry24ns} has challenged most existing binary interaction models. The progenitors of the compact objects must have been AGB or RSG/YSG stars, which are traditionally expected to trigger CEE, forming $\sim$$1$~day period binaries, contrary to the observations. Two viable pathways to form them are: 1) stable mass transfer, where the stability extends to much higher initial mass ratios \citep[][]{Ge20,Temmink23,Yamaguchi25,Bhattacharjee26}, or 2) CEE, but with very efficient envelope ejection potentially with the help of the recombination energy \citep{Yamaguchi24b,Belloni24,Thai26}. The results presented in Section~\ref{sec:mesa_chrom} show that chromospheric mass transfer may impact both evolutionary channels.

Preliminary experiments show that denser chromospheres, which exhibit enhanced envelope stripping, can form low mass He-WD binaries through stable mass transfer at AU-scale periods (unless the initial period is sufficiently long, e.g. case P2250). Such systems do exist in the Gaia sample, but suffer from a negative detection bias owing to low astrometric mass ratio \cite{Shahaf23,Yamaguchi26}. We expect more such systems to be detected in Gaia DR4. The preferred formation channel for the CO-WD binaries then becomes CEE from initially longer periods (e.g., Figure~\ref{fig:gaia_wdms_from_cee}). Lowering the chromospheric densities, such as the reduced chromosphere case, increasingly allows for stable MT channels (e.g., the reduced chromosphere case in P750 where the MT remains stable). Overall, chromospheric mass transfer can alter the mapping of the initial to final parameters of a binary population. The wide WD+MS binaries can therefore help calibrate our model through future population synthesis studies.

%The full chromosphere often lead to substantially early stripping (e.g., case P750) that our model finds it difficult to produce AU-scale CO-core WD+MS binaries with stable mass transfer, unless the initial period is sufficiently long (case P2250). However CEE with efficient envelope ejection can still produce the Gaia WD+MS binaries (Figure~\ref{fig:gaia_wdms_from_cee}).

%Owing to enhanced stripping, for a large fraction of initial periods where the classical prescription forms a CO-core WD, our model forms a low mass He-core WD (case P750 in Section~\ref{sec:mesa_chrom} for example). Thus, our model finds it difficult to produce AU-scale CO-core WD+MS binaries with stable mass transfer, unless the period is sufficiently long (case P2250). We can, however, still form CO-core WD+MS binaries in Gaia DR3-like periods with CEE. We show an example in Figure~\ref{fig:gaia_wdms_from_cee} -- the \citetalias{Ritter88} prescription predicts negligible interaction, whereas the chromospheric mass transfer results in enough initial orbital shrinkage to potentially trigger a CEE. We find that the post-CEE period and component masses well agree with those of the Gaia systems. This shows that CEE remains a viable pathway to produce at least some of the Gaia binaries. A future population-synthetic analysis will enable proper calibration of our model. Additionally, observed distributions of wide binaries with He-core vs CO-core WDs will serve as a further test. 

Chromospheric mass transfer will also impact our understanding of other similar ``wide'' post-interaction systems. Examples include the Blue Lurkers \citep{Leiner19}, where the MS star is spun-up and fast rotating -- a telltale signature of mass transfer history. Previously, WRLOF was used to model the widest systems \citep{Sun24}. Chromospheric mass transfer will enable a similar analysis for a broader set of systems. In many wide binaries, the accretor photospheres appear polluted with carbon and s-process elements (e.g., barium) in the accreted matter from the WD's progenitor AGB star. This often enables measurement of the amount of accreted mass. Recent studies like \cite{Sarmah26} suggest total accreted masses of $\approx$$0.1-0.5~M_{\odot}$. The extended chromosphere can supply such large amounts of transferred matter even for very wide systems where \citetalias{Ritter88}-like prescriptions would have predicted little interaction (e.g., case P2250). In very young post-MT systems, called Abell 35-type systems \citep{Bhattacharjee26}, the companion to the hot WD is both thermally inflated and highly spun-up. A change in the mass transfer mechanism will subsequently influence our interpretation of the accretor's response.

%Another related class is the Abell 35-type systems \citep{Bhattacharjee26}, where the WD is young and the companion is both fast rotating and thermally inflated (thus appearing as a giant) owing to the recent mass transfer episode. Our revised mass transfer mechanism will directly effect the response of the accretor to mass transfer and, thus, interpretation of these systems.

Our model also applies to post-CEE short-period binaries, especially when quantifying the mass transfer that occurred before the onset of the CEE. Particularly surprising are observations suggesting that the companions in post-CEE binaries are often both polluted (for example, post-CEE dwarf carbon stars, \citealt{Roulston21}) and thermally inflated (binaries inside planetary nebulae, \citealt{Jones15,Jones20PN,Jones26}). In fact, \cite{Miszalski13} estimates a total accreted mass of $\approx$$0.05-0.3~M_{\odot}$ for the post-CEE binary inside the Necklace nebula, which is comparable to those estimated for stable mass transfer systems. As the interaction with the CE is expected to be on dynamical timescales, this implies a substantial pre-CEE mass transfer phase. Chromospheric mass transfer naturally accounts for this phase.

\subsection{Symbiotic systems as observational tests}\label{subsec:symbiotic_tests}

Symbiotic systems \citep{Merc19} -- wide binaries where evolved giant donors transfer mass to WD or NS accretors -- are excellent probes of mass transfer physics. The accretion/mass transfer rate can be estimated using various methods, viz. the X-ray/far-UV accretion luminosity \citep[e.g.,][]{Lima24} or the line of sight hydrogen column density \citep[e.g.,][]{Shagatova16}. The radius of the giant donor can be estimated from the optical-infrared SED or spectra \citep[e.g.,][]{Pavlenko16,Skopal17,Sion19}, or directly with interferometric measurements \citep[e.g.,][]{Merc25}. The ellipsoidal variability in light curves and/or radial velocity measurements can provide the orbital periods \citep[e.g.,][]{Mikoajewska03, Nagarajan24}. These measurements, thus, can provide both the Roche fill factor and the corresponding mass transfer rate -- ideal to test our mass transfer model.

These observations have long challenged the classical understanding of mass transfer \citep{Mikoajewska12}. Some of the key puzzles are: 1) In many systems, the donor is significantly under-filling its Roche lobe  ($\sim$$20-60\%$, \citealt{Merc25, Gaudin26}). However, the $\dot{M}$ estimates in some systems \citep[like AG Peg][$\sim 10^{-7\pm1}~M_{\odot}/\rm year$]{Sion19}, or those required to trigger thermonuclear burning on the accreting WD's surface, are too high to be met by \citetalias{Ritter88}-like channels. 2) The inferred mass ratios in many systems are high enough to trigger CEE. The duration of the pre-CEE phase of mass transfer should thus be long enough to be detected. 3) Ellipsoidal variability at the observed amplitude is not expected from Roche under-filled systems. However, the dense chromosphere does fill the Roche lobe and may contribute to the ellipsoidal modulations. In fact, the data for AG Peg is broadly consistent with our model (from Figure~\ref{fig:mdot_1msun_q2_fixedRd}), within the uncertainty limits on $f_{\rm con}$.

%Measurements show that in most systems the donor is significantly under-filling its Roche lobe  ($\sim$$20-60\%$, \citealt{Merc25}). However, mass transfer rate estimates often results in high values of -- e.g., $\sim$$10^{-7\pm1}~M_{\odot}/\rm year$ for AG Peg, \citealt{Sion19}. In general, triggering surface thermonuclear burning on a WD accretor's surface requires episodes of similarly high accretion rates for these systems -- too high for a \citetalias{Ritter88}-like prescription. Additionally, ellipsoidal variability at the observed amplitude is not expected from Roche under-filled systems, unless an extended dense environment fills the Roche lobe. 

%Our model provides a natural explanation for the above. Referring to Figure~\ref{fig:mdot_1msun_q2_fixedRd}, the mass transfer measurements for AG Peg are broadly consistent with our model, within the uncertainty limits on $f_{\rm con}$. More such measurements is the key to observationally calibrate this parameter. The Roche-filling dense chromosphere can also potentially produce the observed ellipsoidal variability. \textbf{A light curve modeling is beyond the scope of this paper (is it?)}. 

X-ray-based accretion rate measurements in symbiotic systems, however, often suggest low values of  $\sim$$10^{-11}~M_{\odot}/\rm year$ \citep{Lima24} -- much lower than suggested by our theory. Though this appears contradictory, it may not be so. With regard to our model, we emphasize that the convective shocks are stochastic. The accretion rates calculated in this work are time-averaged, dominated by the short epochs of high $v_s$ (thus, $\dot{M}$). But at most times, low velocity shocks are expected to prevail. Thus, a low measured $\dot{M}$ for most symbiotic systems with a few outliers (like AG Peg) is not against our model. We predict accretion rates to vary stochastically on time scales of months-years in such systems, occasionally becoming much larger, like those in AG Peg. With regard to observations, the measured accretion rates from X-rays can also be underestimated owing to the accretion disk being optically thick. 

%This apparent contradiction with our theory and other observations like AG Peg can result from several aspects. On the side of observations, 1) this can result from a low $\beta$ and the underlying $\dot{M}$ can be much higher, 2) the accretion disk can be optically thick (indicated by several studies) yielding just a lower limit of the accretion rate. With regards to our theory, we emphasize that the convective shocks are stochastic, but the $\dot{M}$ given by our model equations are time-averaged. This is, thus, dominated by the episodes of high $\dot{M}$ driven by the rare high $v_s$ shocks. But at majority of instances, low $v_s$ shocks are expected to prevail. Thus, a low measured $\dot{M}$ for most symbiotic systems will, in fact, be at par with our model assumptions. 

\subsection{Caveats}\label{subsec:caveats}

The primary caveat in our model is the uncertainty related to the shock properties and the distribution of their speeds. Referring to Equation~\ref{eq:chrom_density_profile}, the density is a very strong function of the assumed shock speed. Thus, the assumed distribution of shock speeds can substantially alter the densities and, consequently, the mass transfer rates. Our assumption of a Gaussian shock-speed distribution and the adopted definition for $v_{\rm con}$ may not be accurate for multiple reasons -- intermittency in turbulence \citep{Hopkins13}, anisotropy \citep{Goldberg22}, contributions from coherent stellar pulsations, or inaccuracy of mixing length theory in predicting convective velocities. See Section~5.6 in \citetalias{Fuller24} for existing discussion. 

The assumptions listed in Section~\ref{subsec:chrom_dens_profile}, most of which have been employed to facilitate an analytic derivation of the Bernoulli constant, add to the model caveats. We expect none of the three assumptions to be strictly valid. 1) The shocks may initially accelerate down a density gradient, but may slow down as they dissipate energy. The shock speed, and thus the shock pressure support, will in fact be radius dependent. 2) The velocity perpendicular to the equipotential surface is expected to be lower than the starting shock velocity, as part of the kinetic energy is expected to go into parallel motion. Both assumptions affect the shock properties and, thus, may translate to substantial uncertainties in the mass transfer rates. 
With regard to calculating the mass transfer rate, our assumption that all the matter contained within the Lagrange potential gets channeled to the accretor may not hold. Mass loss through the outer Lagrange point may drain some of this matter to a circumbinary disk instead.
%Additionally, the assumption of the perpendicularity of the shock velocity to the Roche equipotentials -- used to derive the density profile -- is inconsistent with the shock speed being the mass transfer speed, which demands the shocks to be perpendicular to the nozzle area instead.
These inaccuracies, however, are expected to affect the mass transfer rates less strongly, resulting in uncertainties of factors of a few.

\section{Summary and Conclusions}\label{sec:conclusions}

Understanding the mechanism of mass transfer in binaries is a crucial step in deciphering the formation channels for a variety of post-interaction binaries. Though interaction peaks when the donor's photosphere over-fills its Roche lobe, the onset of mass transfer occurs when the donor is still Roche under-filling -- matter from its outer atmosphere can reach the L1 point and get channeled onto the accretor. The classical model is a thermally supported atmosphere with a very small scale height, which is used in the seminal work of \citet[][or \citetalias{Ritter88}]{Ritter88} to derive the mass transfer rate. However, numerical simulations of giant stars (i.e. AGB/RSG) show that photospheric matter can get pushed out to large distances by shocks, resulting in orders of magnitude denser ``chromospheres''. In this work, we have developed a simple model for mass transfer from these extended stellar chromospheres, which can serve as an extension to the \citetalias{Ritter88} model.

We base our work on the model developed by \citet{Fuller24} where the chromosphere is supported by stochastic convection-driven shocks arising from the stellar photosphere. First, we re-derive the chromospheric density profile as a function of gravitational potential, incorporating the thermal pressure term -- such that at low shock speeds we recover the isothermal atmosphere (Section~\ref{subsec:chrom_dens_profile}). We evaluate it for Roche potentials and calculate the mass transfer rate by assuming that the matter contained between the inner and outer Lagrange potentials is transferred to the accretor (Section~\ref{subsec:chrom_mdot}). We find that, especially for luminous AGB stars and low Roche fill factors, the chromospheric mass transfer rates can be orders of magnitude larger than given by the \citetalias{Ritter88} prescription (Figure~\ref{fig:mdot_1msun_q2_fixedRd}). To estimate the possible effects of this enhanced interaction on binary evolution, we implement our prescription in MESA and perform a few representative MESA runs (Section~\ref{sec:mesa_chrom}). We find that chromospheric mass transfer can lead to enhanced stripping of the donor, much more transferred mass, and substantial changes to the binary outcome (Figure~\ref{fig:quants_v_age}). This may have a significant impact on our understanding of many types of interacting binaries -- AU-scale binaries, carbon and barium stars, systems that underwent pre-common envelope mass transfer, and others (Section~\ref{sec:discussion}).

Our model is simple. It relies on multiple simplifications and assumptions which may not be accurate (Section~\ref{subsec:caveats}). The convective velocities are uncertain, which in turn translate to large uncertainties in the mass transfer rates. Nonetheless, our conclusion that mass transfer rates can be much larger than predicted by the \citetalias{Ritter88} prescription is likely robust. Our model can capture the effects of both the thermal motion and shocks, making it a more general model for mass transfer -- the classical \citetalias{Ritter88} prescription simply becomes a limiting case. Our model is conceptually similar to the empirical Wind Roche Lobe Overflow \citep{Mohamed07,Abate2013} prescription, but more flexible since it is based on analytic models rather than SPH simulations. Appropriate calibration of the chromospheric density with observations and simulations, along with possible improvements to the underlying theory, will enable formulation of a robust theory for binary mass transfer.

% This was used by \cite{Ritter88} to derive a corresponding optically thin mass transfer rate, that is widely used in binary models.

%Thus, a direct prediction of our model is a population of low-mass He-core WD+MS binaries at AU-scale orbits -- in the regime where RGB stripping would not be 

%The nozzle area, $A_N(\phi)$, is non-trivial to calculate. Most previous studies use approximate Taylor-series expansions around the L1 point. But such approximations break down when integrating over large range of potentials such as up till L2/L3. \textbf{Thus, for each $\phi$, we calculate $A_N(\phi)$ numerically.}

% \end{acknowledgements}

\section{Acknowledgments}

We thank the anonymous referee for their comments which improved the manuscript substantially. We thank Dr. Bernd Freytag for kindly sharing the simulation data with us. We acknowledge the use of AI models to debug some of the codes used in this work and generate Figure~1. We have used the \texttt{Python} packages NumPy \citep{harris2020array}, SciPy \citep{2020SciPy-NMeth}, Matplotlib \citep{Hunter:2007}, Pandas \citep{reback2020pandas}, Astropy \citep{Astropy13, Astropy18}, and Astroquery \citep{astroquery19} at various stages of this research. This research was supported in part by NSF grant AST-2540180.

\section{Data Availability}
All the MESA inlists and the relevant codes to reproduce the results of this paper can be found in \url{https://github.com/Soumin1908/Chromospheric-Mass-Transfer}

\appendix

\section{Chromospheric Wind in Single Stars}\label{app:coronal_wind}

\begin{figure}
    \centering
    \includegraphics[width=1\linewidth]{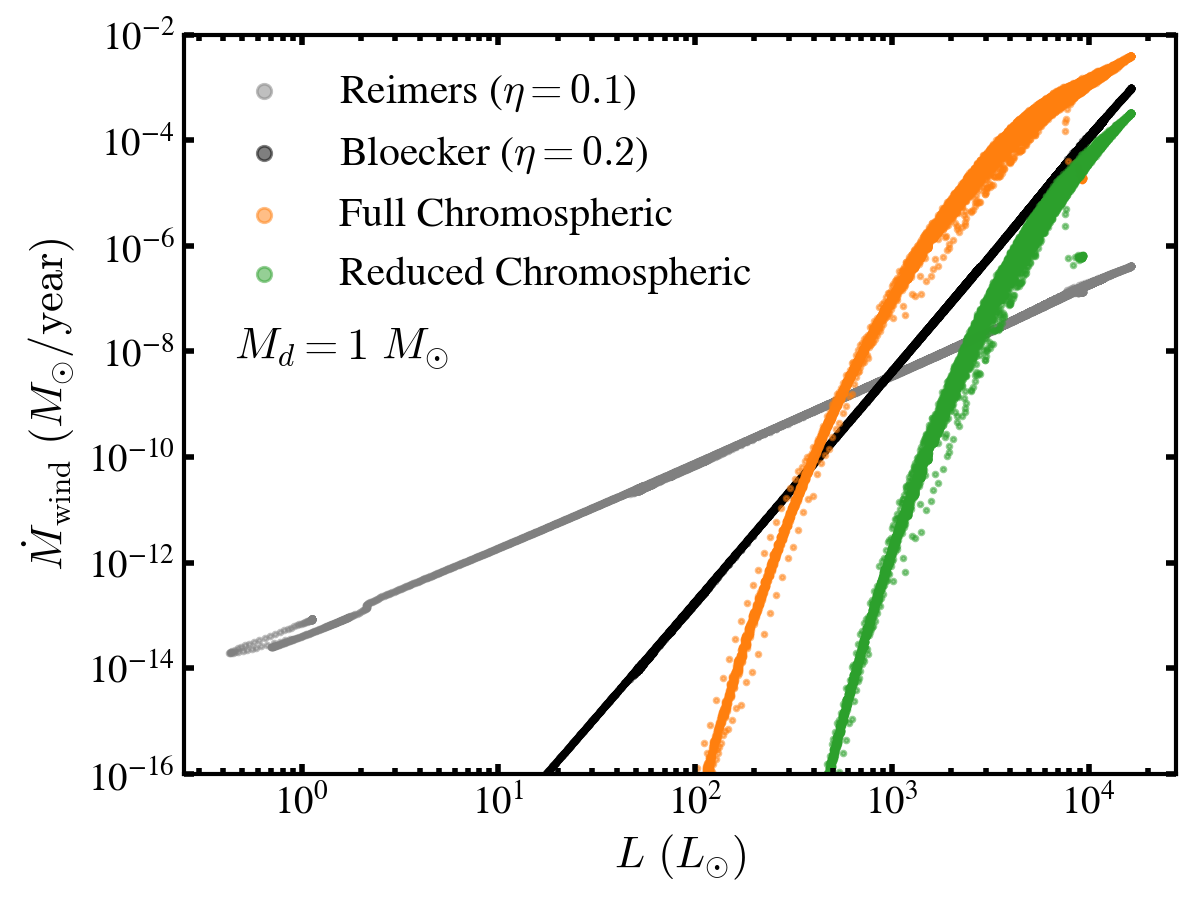}
    \caption{The wind mass loss rate, $\dot{M}_{\rm wind}$, as a function of stellar luminosity, $L$, for a $1~M_{\odot}$ MESA star with three different wind prescriptions: Reimers \citep{Reimers75}, Bloecker \citep{Blocker95}, and Chromospheric wind (Equation~\ref{eq:coronal_wind}) for both full and reduced cases. For the Reimers and Bloecker winds, we follow the same prescription as used in the MESA runs in Section~\ref{sec:mesa_chrom}. To be consistent with \citetalias{Fuller24} we use a dust formation temperature of $T_{\rm dust}=1200$~K.}
    \label{fig:new_wind_comparison}
\end{figure}

The wind mass loss rate consistent with the Chromospheric picture is derived in Section~3.1
in \citetalias{Fuller24}, in the approximation of $v_{\rm tot}=v_s$. Assuming a dust formation temperature of $T_d$,
the dust formation radius is given by:

\begin{equation}
    R_{\rm dust} = R_{\rm ph}\left(\sqrt{1 - \left[1 - 2\left(\frac{T_d}{T_{\mathrm{ph}}}\right)^4\right]^2}\right)^{-1}.
\end{equation}

\noindent The single star wind mass loss rate is then given by mass flow continuity at the dust formation radius:
\begin{equation}\label{eq:coronal_wind}
\begin{aligned}
\dot{M} = 4\pi R_{\rm dust}^2\int_0^{\infty}\rho(R_{\rm dust},\,v_s)\,p(v_s)\,v_{\rm tot}\,dv_s
\end{aligned}
\end{equation}
where $\rho(R_d,\,v_s)$ is given by Equation~\ref{eq:density_profile_single_stars}. This formalism is an improvement from Equation~15 in \citetalias{Fuller24} as it incorporates both the different shock speeds and the thermal velocity. 

\begin{figure}
    \centering
    \includegraphics[width=1\linewidth]{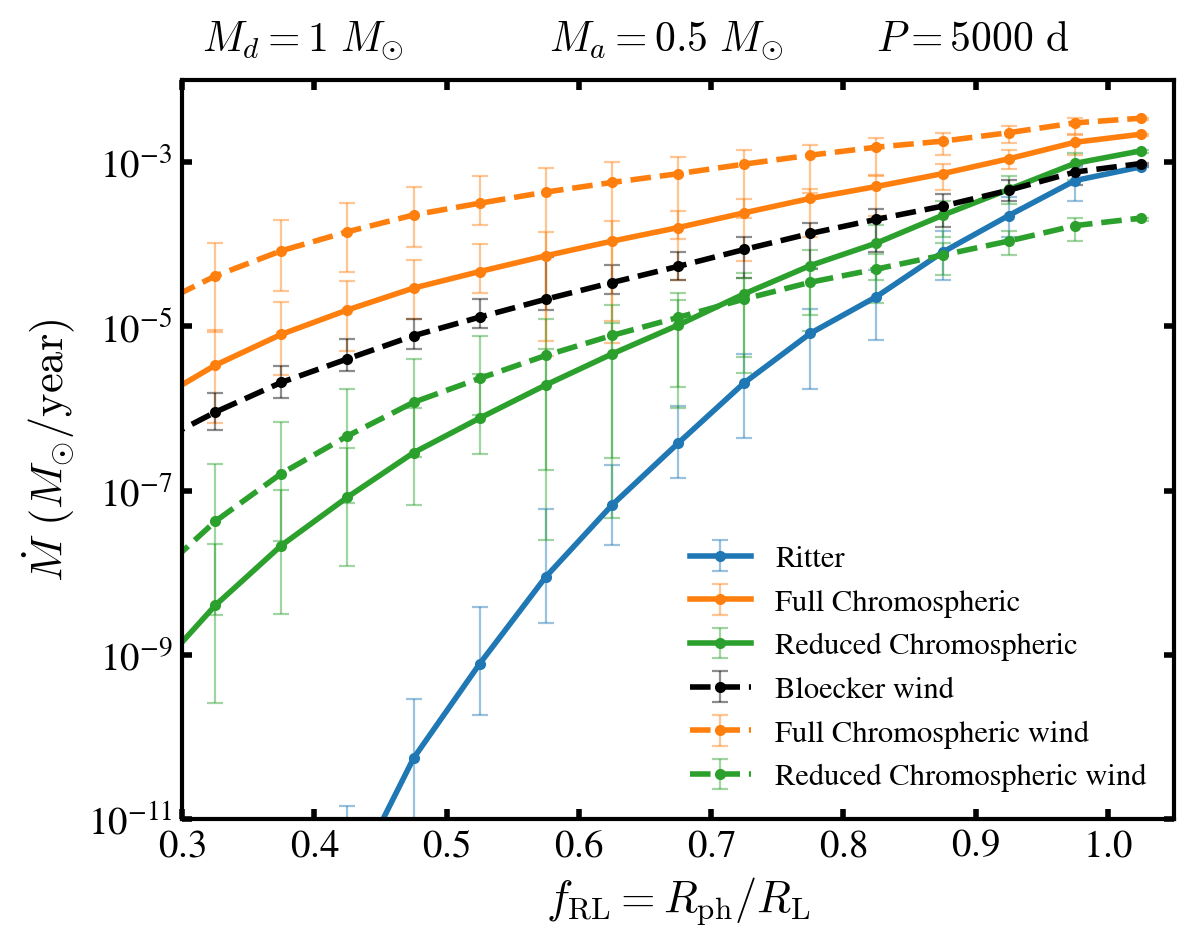}
    \caption{Mass transfer rates (solid lines) compared to the donor's wind mass loss rates (dashed lines) during its AGB phase placed in a fiducial binary with fixed component masses and orbital period. The AGB undergoes thermal pulses and attains a given $f_{\rm RL}$ multiple times with slightly different properties yielding different mass loss rates. The filled circles represent the median of all such instances within $f_{\rm RL}$ bins of $0.05$, and the error bars represent the corresponding range.}
    \label{fig:fixed_period_mass_loss_comp}
\end{figure}

\begin{figure*}
    \centering
    \includegraphics[width=\linewidth]{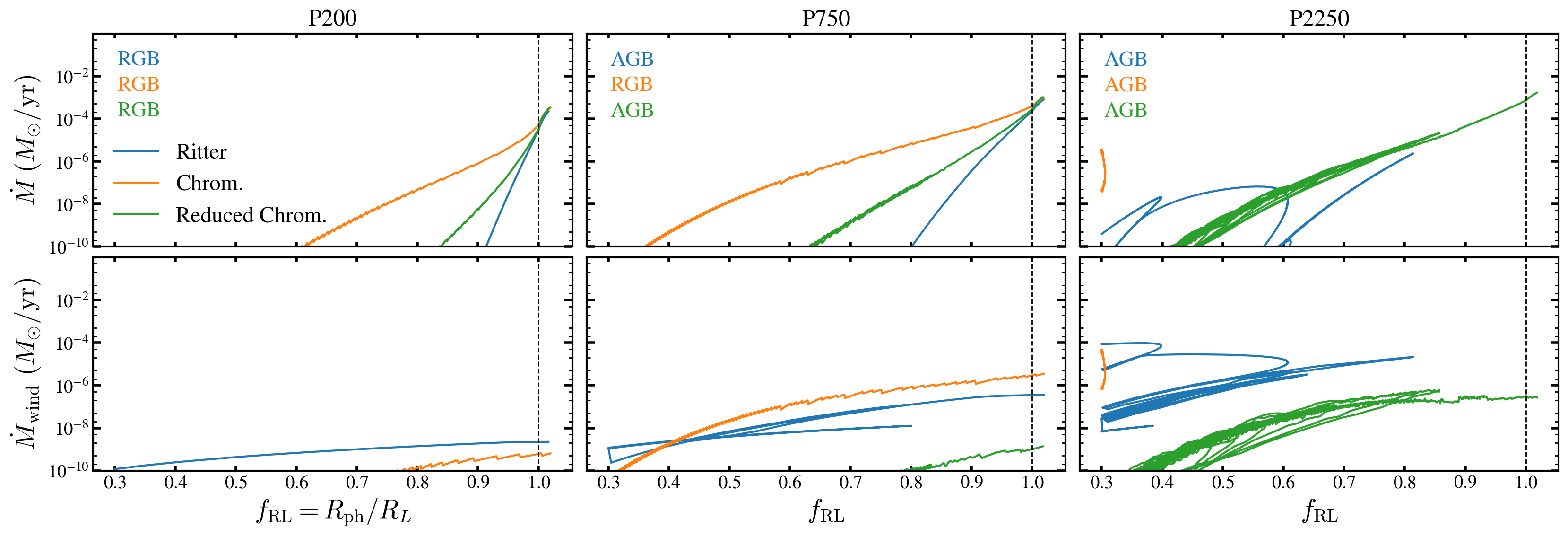}
    \caption{Same as Figure~\ref{fig:quants_v_rdbyrl} but with the Chromospheric wind loss prescription.}
    \label{fig:quants_v_rdbyrl_revised_new_wind}
\end{figure*}

To compare this prescription, which we call the Chromospheric Wind, with the traditional prescriptions of Reimers and Bloecker winds, we calculate the wind mass loss rate, $\dot{M}_{\rm wind}$, for a $1~M_{\odot}$ MESA star at different luminosities during its evolution. We turned off wind during the MESA evolution to keep the mass constant throughout the different phases. The result is shown in Figure~\ref{fig:new_wind_comparison}. Similar to the inferences in \citetalias{Fuller24}, the Chromospheric prescriptions yield a much steeper dependence of $\dot{M}_{\rm wind}$ compared to the other prescriptions. The full chromospheric case predicts very strong winds of up to few orders of magnitudes higher than Bloecker near the tip of the RGB ($L\sim 10^3 \, L_\odot$). The reduced chromosphere, however, predicts weaker winds -- nearing the Bloecker rates during AGB phases. We expect the reality to be somewhere in between.

We then compare the optically thin mass transfer rate to $\dot{M}_{\rm wind}$ for different AGB radii for a fixed binary separation. The result is shown in Figure~\ref{fig:fixed_period_mass_loss_comp}, where we take the same $1~M_{\odot}$ MESA star and put it in a binary at different stages of its AGB phase. We find that in the classical combination of \citetalias{Ritter88} with Bloecker wind, the latter dominates substantially, thus governing the evolution in Roche under-filling evolved donors. However, with the chromosphere, the mass transfer rates are much higher and are comparable to the wind mass loss rates. Thus, even under Roche under-filling conditions, both the channels for donor mass loss affects the evolution of the binary.

\begin{figure*}
    \centering
    \includegraphics[width=1\linewidth]{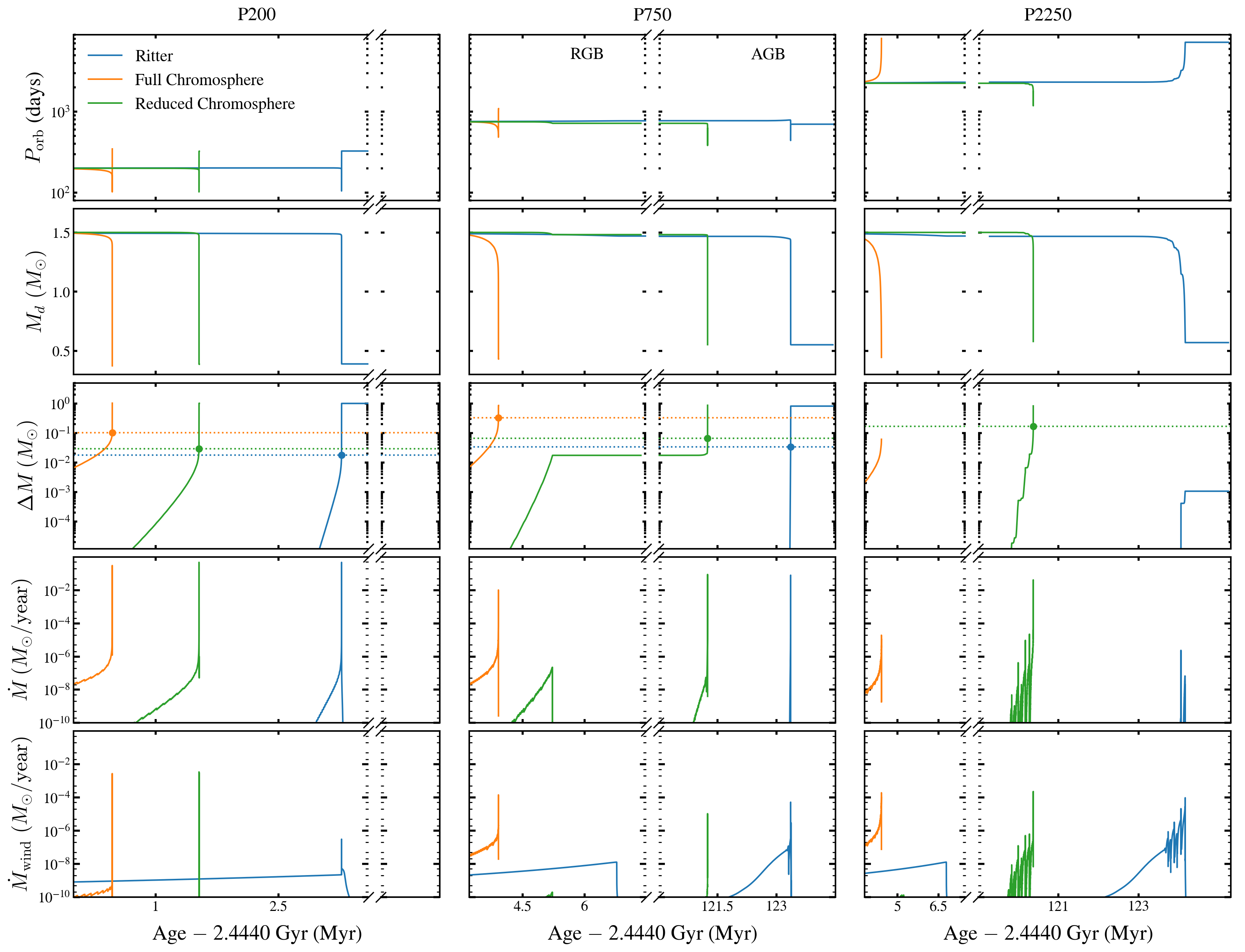}
    \caption{Same as Figure~\ref{fig:quants_v_age} but with the Chromospheric wind loss prescription.}
    \label{fig:quants_v_age_revised_new_wind}
\end{figure*}

We directly implement these prescriptions in MESA to demonstrate the effect of a change in wind prescription. We choose the same initial binary conditions as in Section~\ref{sec:mesa_chrom} -- for the \citetalias{Ritter88} case we retain the wind prescription from the previous section. The results are shown in Figures~\ref{fig:quants_v_rdbyrl_revised_new_wind} and \ref{fig:quants_v_age_revised_new_wind}. The outcomes are broadly similar in the P200 and P750 cases.

The outcome differs significantly in the P2250 case compared to that observed in Section~\ref{sec:mesa_chrom}. The full chromosphere case takes a peculiar turn as the winds become so strong even in the RGB phase that strips off the envelope with only a little mass transferred. Thus, a He-core WD is formed in a widened orbit. The total mass transferred is still about two orders of magnitude larger than the \citetalias{Ritter88} case. In the reduced chromosphere case the star successfully enters the AGB phase. The wind stripping is sufficiently weak such that the star retains enough mass that it enters the optically thick $(f_{\rm RL}>1)$ mass transfer phase. The final outcome is a CO WD-main sequence binary on a $\approx$$2000$~day period. This exercise shows that a change in wind prescription can indeed affect the binary outcome substantially. More in-depth analysis of this aspect will be presented in a later work, where a full binary Chromospheric wind solution will be derived. 

\section{Quantities}\label{app:quants}

Here we list the additional equations and methods used to implement our mass transfer prescription. We use the primed quantities as defined in Section~\ref{subsubsec:chrom_dens_binary}.

The Roche potential is given by:
\begin{equation}\label{eq:roche_pot}
\begin{split}
    -\phi'(x',y',z') = \frac{1}{\sqrt{x'^2+y'^2+z'^2}}+\frac{1}{q\sqrt{(x'-1)^2+y'^2+z'^2}}\\+\frac{1}{2}\left(1+\frac{1}{q}\right)\left[\left(x'-\frac{1}{1+q}\right)^2+y'^2\right],
\end{split}
\end{equation}
where $x'$ is along the binary axis (connecting the two stars, with the donor placed at $x'=0$ and the accretor at $x'=1$) and $z'$ is perpendicular to the binary plane. We solve for the zeros of the $x'$-derivative of Equation~\ref{eq:roche_pot} along $y'=z'=0$ to obtain the $x'_{\rm L1/L2}$ coordinate, and thus the corresponding potentials, at the Lagrange points. 

The \cite{Eggleton83} Roche radius is given by:
\begin{equation}\label{eq:eggleton}
    R'_{L1} = \frac{0.49\,q^{2/3}}{0.6\,q^{2/3} + \ln\left(1 + q^{1/3}\right)}.
\end{equation}

The expressions for the terms appearing in the nozzle area expression (Equation~\ref{eq:dAN_by_dphi}) are:
\begin{equation}
\begin{split}
    C_1' = \frac{1}{2}\left[\frac{1}{|x'_{L1}|^3}+\frac{1}{q|1-x'_{L1}|^3}-(1+1/q)\right]\\
    C_2' = \frac{1}{2}\left[\frac{1}{|x'_{L1}|^3}+\frac{1}{q|1-x'_{L1}|^3}\right]\\
    C_3'=-\frac{3}{8}\left[\frac{1}{|x'_{L1}|^5}+\frac{1}{q|1-x'_{L1}|^5}\right],
    \end{split}
\end{equation}
where $x'_{L1}$ is the location of the L1 point along the binary axis, which is obtained from Equation~\ref{eq:roche_pot}.

\section{Approximations for Numerical Areas}\label{app:numerical_area}

\begin{figure*}
    \centering
    \includegraphics[width=1\linewidth]{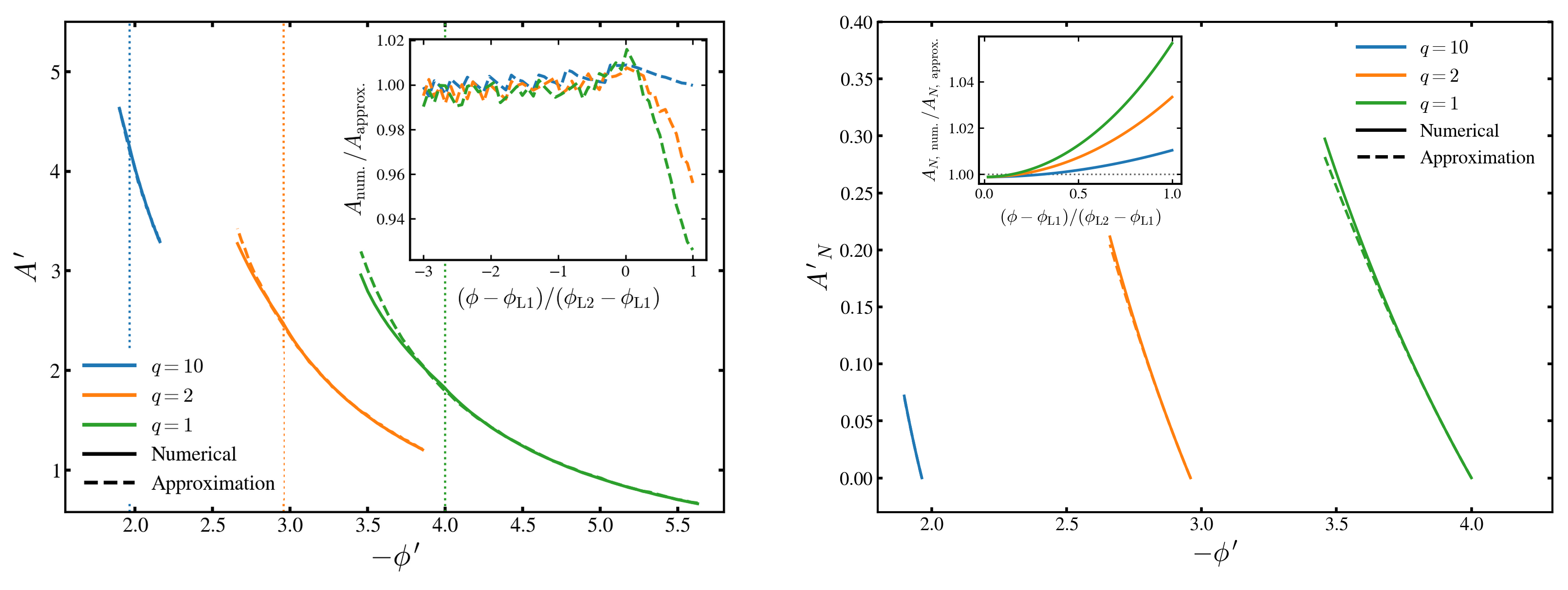}
    \caption{Comparing the numerical areas (solid line) and the corresponding approximations (dashed lines) for the Roche equipotential surface (left) and nozzle cross section (right). The insets show the ratio of the two quantities. We only show results for $q>1$, which is the primary regime of our interest, but a similar comparison holds for practically all mass ratios. We plot up to $\phi_{\rm L2}$, the maximum potential of interest for our purpose.}
    \label{fig:numerical_vs_approx_area_comp}
\end{figure*}

The two primary (semi-)analytic approximations to numerical areas that enter into our expressions in Sections~\ref{subsubsec:chrom_dens_binary} and \ref{subsec:chrom_mdot} are the Roche equipotential surface areas and the nozzle areas. To approximate the former, we use the surface area of the volume-equivalent spheres defined by Equation~\ref{eq:vol_eq_pot}. For the latter, we use the relation provided in \cite[][Equation~\ref{eq:dAN_by_dphi}]{Jackson17}. Here we present the comparison with the numerically evaluated values.

We employ grid-based numerical integration for both the quantities. For the equipotential surface for a given potential, $\phi$, we first solve for the two ends of the surface along the $x$-axis using Equation~\ref{eq:roche_pot}. Then, for each $x$-grid point within the boundaries of the surface, we traverse an angular grid in the $y-z$ plane to solve for the corresponding ``radial'' coordinates on the equipotential surface. We call this a ring. We then evaluate the area between two consecutive rings by summing up the areas of the differential triangles formed by the rings' coordinates. The total surface area is the sum of all the consecutive ring areas. The left panel of Figure~\ref{fig:numerical_vs_approx_area_comp} shows the comparison between the numerical and the spherical-approximate areas, in primed quantities for three mass ratios. They agree very well, within $\lesssim$$10\%$, which is well within the error margins of our model. 

The numerical nozzle areas were calculated in a similar way -- solving for points on a ``ring'' around the L1 point and summing up the differential areas, this time on the face of the ring. This yields the cross-sectional area of a given potential surrounding the L1 point. The right panel of Figure~\ref{fig:numerical_vs_approx_area_comp} shows the comparison with the approximate area obtained by integrating Equation~\ref{eq:dAN_by_dphi} up to $\phi_{\rm L2}$. They agree within $\lesssim$$5\%$.

\section{Reduced Formula for Chromospheric $\dot{M}$}\label{app:reduced_chrom_mdot}

Equation~\ref{eq:mdot_chrom_condensed} can be expanded as:
\begin{equation}
\begin{split}
    \dot{M}(v_s) = \sqrt{\frac{2}{\pi e}}\frac{\rho_{\rm ph}}{v_{\rm con}}GM_da \\\times\int_0^{\infty}\int_{\phi_{L1}'}^{\phi_{\rm max}'}e^{-\frac{\max\left(\phi'-\phi_{\rm ph}'^V,~0\right)}{v_{\rm tot}'^2}}\left(\frac{f_{\rm RL}R'_{\rm L1}}{R'^V_{\phi}}\right)^{\frac{2v_s'^2}{v'^2_{\rm tot}}}\\\times e^{-\frac{v_s'^2}{2v'^2_{\rm con}}}\frac{dA'_N(\phi')}{d\phi'}\,v_{\rm tot}'dv_s'\,d\phi',
\end{split}
\end{equation}
where the modified density exponential accounts for the saturated regime for mass transfer. The complexity of this expression can be significantly reduced by computing an approximate expression for the integral over $\phi$. For any mass ratio, $\phi'_{\rm max}$ is within $\lesssim$$15\%$ of $\phi'_{L1}$. Over this range of potentials, $R_{\phi}'^V$ varies at most by a factor of $\sim$$2$ and $dA'_N(\phi')/d\phi'$ varies by $\approx$$20\%$. Thus, the dominant contribution comes from the exponential. We thus define an empirical mean potential:
\begin{equation}
    \phi'_m=\frac{\phi'_{L1}+\phi'_{\rm max}}{2}.
\end{equation}
We then evaluate the corresponding $R_{\phi'_m}'^V$ and $|dA'_N(\phi')/d\phi'|_{\phi'=\phi'_m}$ and keep them fixed. With this simplification, we perform the integral on $\phi'$ analytically. This approximate form agrees with the true integral within a few tens of percent in almost the whole range of parameter space. The final expression for the mass transfer rate thus becomes:
\begin{equation}\label{eq:mdot_rescaled_I}
    \dot{M}(v_s) = \sqrt{\frac{2}{\pi e}}\frac{\rho_{\rm ph}}{v_{\rm con}}GM_da\times\mathcal{I}_{\rm red}(q, f_{\rm RL}, v_{\rm con}', c_T'),
\end{equation}
where 
\begin{equation}\label{eq:I_red}
\begin{split}
    \mathcal{I}_{\rm red}(q, f_{\rm RL}, v_{\rm con}', c_T') \\= \left|\frac{dA'_N(\phi')}{d\phi'}\right|_{\phi'_m}\int_0^{\infty}\left(\frac{f_{\rm RL}R'_{\rm L1}}{R'^V_{\phi'_m}}\right)^{\frac{2v_s'^2}{v'^2_{\rm tot}}}e^{-\frac{v_s'^2}{2v'^2_{\rm con}}}\\\times f(\phi'^V_{\rm ph},~\phi'_{L1},~\phi'_{\rm max},~v_{\rm tot})\,v_{\rm tot}'dv_s',
\end{split}
\end{equation}
and
\begin{equation}\label{eq:f_phi_vrot}
\begin{split}
    f(\phi'^V_{\rm ph},~\phi'_{L1},~\phi'_{\rm max},~v_{\rm tot}')\\=\max(\phi'_{\rm ph}-\phi_{L1},~0)-\max(\phi'_{\rm ph}-\phi_{\rm max},~0)\\+v_{\rm tot}^2\left(e^{-\frac{\max\left(\phi'_{L1}-\phi_{\rm ph}'^V,~0\right)}{v_{\rm tot}'^2}} - e^{-\frac{\max\left(\phi'_{\rm max}-\phi_{\rm ph}'^V,~0\right)}{v_{\rm tot}'^2}}\right).
\end{split}
\end{equation}
Thus, Equations~\ref{eq:mdot_rescaled_I}, \ref{eq:I_red}, and \ref{eq:f_phi_vrot} together constitute our final set of equations that we use to calculate the mass transfer rate both for Figure~\ref{fig:mdot_1msun_q2_fixedRd} and Section~\ref{sec:mesa_chrom}. In the limit of $v_{\rm con}\rightarrow0$, our model approaches the \citetalias{Ritter88} rate.

\section{Revised implementation of R88 prescription in MESA}\label{app:mesa_bonk}

\begin{figure}[t]
    \centering
    \includegraphics[width=1\linewidth]{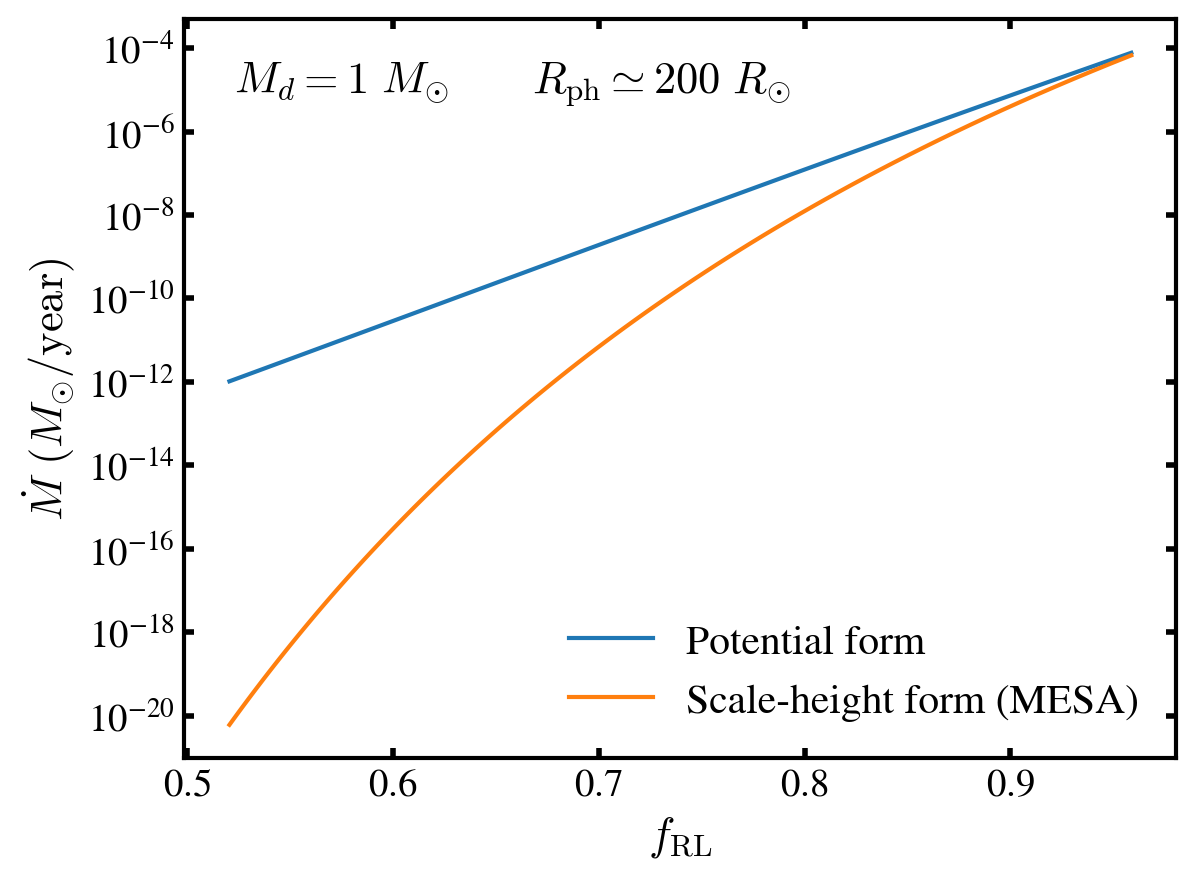}
    \caption{Comparing the \citetalias{Ritter88} mass transfer rate using the density profiles given by Equation~\ref{eq:rho_ritter} (blue, general form) and Equation~\ref{eq:rho_ritter_radius} (orange, approximate form and current MESA implementation). They converge at large $f_{\rm RL}$, but, at low $f_{\rm RL}$, the current implementation underestimates the $\dot{M}$ by orders of magnitude. }
    \label{fig:ritter_comp_radvpot}
\end{figure}

MESA currently implements the \citetalias{Ritter88} prescription for optically thin mass transfer by using Equation~9 in their paper, which uses an expression for density in terms of radius and an effective scale height ($H_p$, Equation~8a) in their paper:
\begin{equation}\label{eq:rho_ritter_radius}
    \rho=\frac{1}{\sqrt{e}}\rho_{\rm ph}e^{-\frac{R_{L1}-R_{\rm ph}}{H_p}},
\end{equation}
where 
\begin{equation}
    H_p=\frac{P_{\rm ph}}{g\rho_{\rm ph}}\approx\frac{c_T^2R_{\rm ph}^2}{GM_d}.
\end{equation}
However, note that the more fundamentally correct expression for the density is given by Equation~\ref{eq:rho_ritter} (same as their equation~3). A crucial assumption going into obtaining Equation~\ref{eq:rho_ritter_radius} from Equation~\ref{eq:rho_ritter} is $R_{L1}\approx R_{\rm ph}$, which breaks down for low Roche fill factors. This leads to gross underestimation of the mass transfer rate in this regime. This can readily be seen by considering a simple spherical potential, where $\phi^S=GM_d/R$. Then, we have:
\begin{equation}
    \frac{R_{L1}-R_{\rm ph}}{H_p}=\frac{R_{L1}}{R_{\rm ph}}\frac{\phi^S_{L1}-\phi^S_{\rm ph}}{c_T^2}>\frac{\phi_{L1}-\phi_{\rm ph}}{c_T^2}.
\end{equation}
This shows that the density, and consequently the mass transfer rate, estimated using Equation~\ref{eq:rho_ritter_radius} (which MESA currently implements) will be exponentially smaller than that calculated using Equation~\ref{eq:rho_ritter}. We show one example of this discrepancy in Figure~\ref{fig:ritter_comp_radvpot}. The two methods of calculating $\dot{M}$ agree reasonably well for $R_d/R_{\rm L1}\gtrsim0.85$, but they differ by orders of magnitude for lower Roche fill factors. 

We do not expect this to affect most binary simulations, especially where the majority of mass transfer occurs during the optically thick phase. But there can be significant differences when considering binary separations where interaction is taking place during low Roche fill factors of the donors (unevolved donors or wide binaries). For example, the default implementation can significantly overestimate the required Roche filling factor when comparing to observations of systems undergoing optically thin mass transfer (e.g. symbiotic systems). Overall, the amount of mass transferred during the optically thin phase can be substantially underestimated. For example, in the P2250 run in our work (with the \citealt{Ritter88} prescription), the default implementation yields $\Delta M\approx8\times10^{-6}~M_{\odot}$, whereas the revised implementation yields two orders of magnitude higher value of $\Delta M\approx1\times10^{-3}~M_{\odot}$.

\section{Post-CEE period calculation}\label{app:post_cee}

We adopt the $\alpha$-prescription for CEE as used in \cite{Yamaguchi24b}. The CE binding energy, $E_{\rm bind}$, is related to the initial and final orbital energies as:
\begin{equation}
    E_{\rm bind} = \alpha_{\rm CE}\left(-\frac{GM_{\rm WD}M_{a,f}}{2a_f}+\frac{GM_{d,i}M_{a,i}}{2a_i}\right),
\end{equation}
where $M_{\rm WD}$ is the WD mass, which is equal to the core mass of the donor. $M_a$ and $M_d$ denote the accretor and donor masses, respectively, with the subscripts $i$ and $f$ indicating the initial (at the onset of CEE) and final masses, respectively. We assume that no mass is accreted during the CEE; thus we set $M_{a,i}=M_{a,f}$. The parameter $\alpha_{\rm CE}$ determines the fraction of orbital energy that goes into unbinding the envelope. We set it to a fiducial value of $\alpha_{\rm CE}=0.3$. The envelope binding energy comprises three energy components: the gravitational binding energy $E_{\rm grav}$, the thermal energy $E_{\rm th}$, and the recombination energy, $E_{\rm rec}$. The total energy then becomes:
\begin{equation}
    E_{\rm bind} = E_{\rm grav}+\alpha_{\rm th}E_{\rm th}+\alpha_{\rm rec}E_{\rm rec},
\end{equation}
where the factors $\alpha_{\rm th}$ and $\alpha_{\rm rec}$ govern the fraction of these energies that go into unbinding the envelope (the rest being radiated away). We use $\alpha_{\rm th}=1$. We then calculate the final periods using both $\alpha_{\rm rec}=0$ and $1$.

\bibliography{ref}{}
\bibliographystyle{aasjournal}

%% This command is needed to show the entire author+affiliation list when
%% the collaboration and author truncation commands are used.  It has to
%% go at the end of the manuscript.
%\allauthors

%% Include this line if you are using the \added, \replaced, \deleted
%% commands to see a summary list of all changes at the end of the article.
%\listofchanges

\end{document}